\documentclass[preprint2,longabstract,natbib,8pt]{aastex}

\usepackage{graphicx}
\usepackage{color}
\usepackage{rededits}
\usepackage{lscape}
\usepackage{multirow}
\slugcomment{}

\shorttitle{lightning keeps MRI}
\shortauthors{Muranushi}

\noproofreadmark

\def\ISSS{IS05}
\def\HGB{HGB95}
\def\SMUN{SMUN00}
\def\ZUMIZUMI{O09}

\newcommand{\simleq}
{\mbox{\raisebox{-0.5ex}{$\textstyle \sim$}
 \raisebox{ 0.8ex}{$\textstyle  \!\!\!\!\!\! <$  }}}

\def\erf{{\mathrm {erf}}}

\def\alfvenVelocityZ{v_{\mathrm {{}_{AZ}}}}

\def\attenuationLengthCR{\chi_{\mathrm {CR}}}
\def\AU{{\mathrm {AU}}}
\def\BoltzmannConstant{{k_B}}
\def\dpCoef{f_\mathrm{DP}}
\def\cdAbove{\chi_{{}_\top}}
\def\cdBelow{\chi_{{}_\bot}}
\def\criticalCurrent{J_{\mathrm {crit}}}
\def\criticalEfield{{E'}_{\mathrm {crit}}}
\def\current{J}
\def\density{\rho}
\def\dustChargeZSq{{\langle Z^2 \rangle}}
\def\dustCrossSection{\sigma_d}
\def\dustGasRatio{f_d}
\def\dustMass{{m_d}}
\def\dustMaterialDensity{\rho_s}
\def\dustNumberDensity{n_d}
\def\dustRadius{a_d}

\def\Ecom{E'}
\def\electronNumberDensity{n_e}
\def\electronThermalVelocity{v_e}
\def\ElementaryCharge{e}
\def\fillingFactor{f_{\mathrm {fill}}}
\def\lhs{{Q_{\mathrm{whb}}}}
\def\saturationJF{f_{\mathrm{sat}}}
\def\saturationJFValue{10}
\def\sustainFactor{f_{\mathrm {whb}}}

\def\fiducialMolecularWeight{\mu}
\def\ionNumberDensity{n_i}
\def\ionizationEnergy{\Delta W}
\def\ionizationRate{\zeta}
\def\ionizationRateCR{\zeta_{\mathrm {CR}}}
\def\ionizationRateRA{\zeta_{\mathrm {RA}}}
\def\jouleHeating{W_{\mathrm{J}}}
\def\gasMass{{\mu m_H}}
\def\gasNumberDensity{n_g}
\def\GravitationalConstant{{G}}
\def\lambdaRes{\lambda_{\mathrm {res}}}

\def\magnet{B}
\def\magneticReynolds{R_{\mathrm{M}}}
\def\MassOfElectron{{m_e}}
\def\MassOfHydrogen{{m_H}}

\def\orbitalOmega{\Omega}
\def\pressure{P}
\def\magDiff{\eta}
\def\magDiffD{{\magDiff_d}}
\def\magDiffE{{\magDiff_e}}
\def\magDiffI{{\magDiff_i}}
\def\rOverAU{{\left(\frac{r}{\AU}\right)}}
\def\scaleHeight{H}
\def\shearWork{W_{\mathrm {sh}}}
\def\sigmaVd{{\langle\sigma v\rangle_d}}
\def\sigmaVe{{\langle\sigma v\rangle_e}}
\def\sigmaVi{{\langle\sigma v\rangle_i}}
\def\soundSpeed{c_s}
\def\SpeedOfLight{c}
\def\starMass{M_{*}}
\def\stickingE{s_e}
\def\stickingI{s_i}
\def\stressTensor{\langle w_{xy} \rangle}
\def\surfaceDensity{\Sigma}
\def\temperature{T}
\newcommand{\unit}[1]{\ \mathrm{#1}}
\newcommand{\Unit}[1]{\mathrm{#1}}

\def\equipartitionCurrent{{\current_{\mathrm {eqp}}}}

\def\equipartitionMagnet{\magnet_{\mathrm {eqp}}}
\def\fiducialSurfaceDensity{{\surfaceDensity_0}}
\def\surfaceDensityFactor{{f_\surfaceDensity}}
\def\fiducialSurfaceDensityPower{q}
\def\fiducialTemperature{\temperature_0}
\def\fiducialTemperaturePower{{\frac 1 2}}
\def\linearMagDiff{\magDiff_0}

\begin{document}

\title{Interdependence of Electric Discharge and The Magnetorotational Instability in Protoplanetary Disks}


\author{Takayuki Muranushi}
\affil{The Hakubi Center for Advanced Research, Kyoto University,
  Sakyo-ku, Kyoto, 606-8502, Japan; muranushi.takayuki.3r@kyoto-u.ac.jp}

\author{Satoshi Okuzumi}
\affil{Nagoya University, Furo-cho, Chikusa-ku, Nagoya, 464-8601, Japan; okuzumi@nagoya-u.ac.jp}

\author{Shu-ichiro Inutsuka}
\affil{Nagoya University, Furo-cho, Chikusa-ku, Nagoya, 464-8601, Japan; inutsuka@nagoya-u.ac.jp}



\begin{abstract}

We study how the magnetorotational instability (MRI) 
in protoplanetary disks is affected by 
the electric discharge caused by the electric field in the resistive MHD.
We have performed
 three-dimensional shearing box simulations with 
various values of plasma beta 
and electrical breakdown models.
\addspan{
We find the self-sustainment of the MRI in spite of the high resistivity.
The instability gives rise to the large electric field that causes the electrical breakdown,
and the breakdown maintains the high ionization degree required for the instability.
}
The condition for this self-sustained MRI is set by
the balance between the energy supply from the \addspan{shearing motion} and
the energy consumed by the Ohmic dissipation.
We apply the condition to various disk models and study where
the active, self-sustained, and dead zones of MRI are. 
In the fiducial minimum-mass solar nebula (MMSN) model,
 the newly-found sustained zone occupies only the limited volume of the disk.
In the late-phase gas-depleted disk models, however,
 the sustained zone occupies larger volume of the
disk.

\end{abstract}





\section{Introduction} \label{introduction}

\addspan{
Protoplanetary disks are the sites of planet formation. 
The disk turbulence greatly affects
the mutual
sticking of the planetesimals, their settlement to the disk midplane. The turbulence is the
source of the angular momentum transfer in the disk that causes
gas accretion and migration of the planetesimal onto the central star. 
Thus understanding the
evolution of the turbulence within protoplanetary disks is an
essential step both in the studies of the disk evolution and the
planet formation.
}

The magnetorotational instability (MRI) is considered to be the major
source of turbulence in many types of accretion disks including
protoplanetary disks (\citet{balbus_instability_1998} and references
therein).  One of the distinct properties of the protoplanetary disks
compared to other accretion disks is that the major parts of the
protoplanetary disks are only weakly ionized, and the magnetic
diffusivity affects the MRI
(\citet{sano_saturation_1998,fleming_effect_2000}.) The low ionization
degree is due to their low temperature and high number density of the
dust component.

In protoplanetary disks, MRI and the dust components affect each other.
The turbulence is one of the source for the relative velocities of the colliding dust
\citep{ormel_closed-form_2007,brauer_planetesimal_2008}, and contributes both on dust growth and disruption
\citep{blum_growth_2008,wada_numerical_2008,guettler_outcome_2010,wettlaufer_accretion_2010}.
On the other hand, dust particles in protoplanetary disks are the major sites of
charged particle recombination, and thereby influences the
ionization degree of the disk.
\citep{sodha_charging_2009, 
  umebayashi_effects_2009, grach_flow_2010}

The dead zone can occupy a large volume of a protoplanetary disk,
especially in the presence of abundant small dust grains
\citep{gammie_layered_1996, sano_magnetorotational_2000,ilgner_ionisation_2006}. However,
various electric discharge mechanisms in protoplanetary disks have been
proposed \citep{horanyi_chondrule_1995, desch_generation_2000,
  muranushi_dust-dust_2010} which may provide higher ionization degree
compared to the values predicted by the dust-absorption equilibrium
models, resulting in the increased MRI activity in the disk.
\addspan{
They consider the electron avalanche process, an exponential growth in
the number of conducting electrons that takes place when the kinetic
energy of the electrons exceeds the ionization energy of a neutral gas
molecule.  The result is electrostatic breakdown, the lowering of the
resistivity of the fluid and electric discharge, increase of the
electric current through the fluid. 
}

\addspan{ Moreover, a model is proposed where the MRI itself provides
  sufficient ionization \citep[hereafter
    \ISSS]{inutsuka_self-sustained_2005}.
  \ISSS\ have shown that
  that the
  electric field typically generated by the protoplanetary disk
  turbulence is strong enough to drive the electrons away from the
  thermal Maxwell-Boltzmann distribution.  Those energetic electrons
  contained in electric current cause the electric discharge and
  maintains the ionization degree high enough for the MRI to survive.
  \ISSS\ have also shown that the energy supply from the shearing motion is
  about 30 times larger than the energy required to 
  maintain the sufficient number of electrons in the presence of
  standard dust grains.
}

However, \ISSS\ have studied only
one-zone models, and with only one set of parameters typical to $~1
\unit{AU}$ of the disk.  In this work, we extend the model IS05 to a
local, 3D simulations of protoplanetary disks and study the
interaction of the MRI with the discharge ionization.  We also apply
the model to global models of protoplanetary disks and study where and
when in the disk the self-sustainment of MRI takes place.

This paper is organized as follows. in \S 2, we perform the numerical simulations of the MRI in unstratified,
three-dimensional shearing-boxes along the lines of \citet[hereafter
  \HGB]{hawley_local_1995} with the nonlinear Ohmic diffusivity added. 
In \S 3, We analyze the activity of the MRI in the protoplanetary disk, using
the method of \citet[hereafter \SMUN]{sano_magnetorotational_2000}.
To calculate the ionization degree in the disk we use the
method proposed by \citet[hereafter \ZUMIZUMI]{okuzumi_electric_2009}.
\S 4 is devoted to conclusions and discussions.
Table \ref{tblNotation} lists the symbols frequently used in this paper.


\begin{table*}
  \begin{center}
    \begin{tabular}{lclc}
      \hline      
      \hline
      \multicolumn{1}{c}{Symbol} &
      \multicolumn{1}{c}{Value (Dimension)} &
      \multicolumn{1}{c}{Definition} &
      \multicolumn{1}{c}{Location} \\
      \hline
      $\density$  & $(\Unit{g\ cm^{-3}})$ & Gas density  & (\ref{basicEqDensity})  \\
      $\mathbf v$ & $(\Unit{cm\ s^{-1}})$ & Gas velocity & (\ref{basicEqVelocity}) \\
      $\mathbf B$ & $(\Unit{g^{1/2}\ cm^{-1/2}\ s^{-1}})$ & Magnetic field & (\ref{basicEqMagnet}) \\
      $P$ &  $\soundSpeed^2 \rho$  & Pressure with isothermal equation of state & (\ref{isothermalEqOS}) \\
      ${\mathbf E}$ & $(\Unit{g^{1/2}\ cm^{-1/2}\ s^{-1}})$ & Electric field in the lab frame  & (\ref{basicEqEfield}) \\
      ${\mathbf \Ecom}$ & $(\Unit{g^{1/2}\ cm^{-1/2}\ s^{-1}})$ & \addspan{Electric field in the comoving frame} & (\ref{basicEqEfieldLab}) \\
      ${\mathbf J}$ & $(\Unit{g^{1/2}\ cm^{-3/2}\ s^{-1}})$ & Electric current & (\ref{basicEqCurrent}) \\
      $\linearMagDiff$ & $(\Unit{cm^2\ s^{-1}})$ & Linear coefficient and --- \\
      $\criticalCurrent$ &  $(\Unit{g^{1/2}\ cm^{-3/2}\ s^{-1}})$ & --- critical current for extended Ohm's law & (\ref{eqNonlinearMagDiff})\\
      ${P_{0}}$ & $(\Unit{g\ cm^{-1}} s^{-2})$ & Initial pressure \\
      $\beta$  &  ${2 \soundSpeed^2}/{\alfvenVelocityZ}^{2}$ & Plasma beta &  \\
      ${B_{z0}}$ & $\sqrt{\beta}\equipartitionMagnet$ & Initial, vertical net magnetic field & (\ref{eqPlasmaBeta})\\
      $\scaleHeight$& $\soundSpeed / \orbitalOmega $ & Disk scale-height & (\ref{eqScaleHeight}) \\
      $\equipartitionMagnet$ & $\sqrt{8 \pi P_0}$           & The nondimensionalization unit of magnetic field &
      (\ref{eqUnitMagnet})\\
      $\equipartitionCurrent$ & $c \equipartitionMagnet / 4\pi H$ & The nondimensionalization unit of current &
      (\ref{eqUnitCurrent})\\
      $\magneticReynolds$ & $ {\alfvenVelocityZ}^2 / \magDiff_0 \Omega $ &
      Magnetic Reynolds number & \S \ref{sectionSimulations}\\
      \\
      $\surfaceDensityFactor$ & $1.0$        & Surface Density Multiplier & (\ref{eqSurfaceDensityModel})\\
      $\fiducialSurfaceDensityPower$ & $3/2$ & Power Law Index of the Surface Density & (\ref{eqSurfaceDensityModel})\\
      $\dustRadius$  & $0.1 \unit{\mu\ m}$   & Radius of Solid Dust Particle & (\ref{eqDustMass})\\
      $\dustGasRatio$  & $0.01$              & Dust to Gas Ratio & (\ref{eqDustNumberDensity})\\
      \hline
    \end{tabular}
  \end{center}
  \caption{The list of symbols used in this paper.}\label{tblNotation}
\end{table*}

\section{Simulations of The MRI with nonlinear Ohm's law}

\subsection{Numerical Setup}\label{sectionNumericalSetup}

\addspan{ There are three diffusion terms in MHD; they are Ohmic
  diffusion, Hall diffusion and ambipolar diffusion. In protoplanetary
  disks any one of the three modes can be the dominant mode depending
  on dust and gas density \citep[e.g.][]{wardle_magnetic_2007}, and
  interaction between the different modes may alter the MRI
  \citep{wardle_hall_2012}.  In this paper we only focus on the Ohmic
  diffusion because it is the most studied one in the context of the
  MRI.  We leave the treatment of other diffusion modes for future
  studies.}

The electric discharge taken into account,
we construct a simple model of the discharge as follows, in terms of an appropriate $\linearMagDiff$ and 
$\criticalCurrent$:
\begin{eqnarray}
  {\mathbf \Ecom} &=& \frac{4\pi}{c^2}  \magDiff\left(\current\right){\mathbf \current}, 
  \label{eqNonlinearEfield} \\
  \magDiff\left(\current\right) &=& \ \ \ \ \ \ \linearMagDiff \ \ \ \ \mathrm {if} \ \ \ \ \current < \criticalCurrent, \nonumber \\
  &=& \frac{\criticalCurrent}{\current}\linearMagDiff   \ \ \ \ \mathrm {if} \ \ \ \ \current > \criticalCurrent.
  \label{eqNonlinearMagDiff}
\end{eqnarray}
This nonlinear diffusivity model states that the electric field on the fluid co-moving frame never exceeds 
a critical value, $ \criticalEfield = 4 \pi c^{-2}\linearMagDiff \criticalCurrent $, thus the magnetic diffusivity 
$\magDiff$ varies depending on electric current $\current$, and Ohm's law become nonlinear.

Note that the smallest space scale dealt in this paper is of order of
$10^{-2}\unit{AU}$. The actual scale
of the discharge structures can be much smaller than this.
The estimate for the lower limit of the
size of such structures is their thickness, which is of the order of
$5000$ times electron mean free path \citep{pilipp_large_1992}.

\addspan{
However, we can derive the macroscopic discharge model
from
the microscopic discharge relation $|\Ecom| < \criticalEfield$ that holds
everywhere in the plasma.
Since e.g. the $x$-component of the discretized electric field
$\langle {\Ecom}_x \rangle$ is obtained from the line integral of the real field over the discretization
length $\Delta V$,}
\addspan{
  \begin{eqnarray}
    |\langle {\Ecom}_x \rangle| &=& 
    \left| \frac{\int {\Ecom}_x  \mathit{dV}}{\Delta V}\right|
    \nonumber \\
    &\leq& \frac{\int |{\Ecom}_x|  \mathit{dV}}{\Delta V}
    \nonumber \\
    &\leq& \frac{\int \criticalEfield   \mathit{dV}}{\Delta V}
    \nonumber \\
    &=&\criticalEfield .
  \end{eqnarray}
}
\addspan{
Therefore, we can use Eqs. (\ref{eqNonlinearEfield}), (\ref{eqNonlinearMagDiff}) 
as a ``coarse-grained model''
where we can interpret $\Ecom$, $J$ as spatial averages. 
If electrical breakdown occurs in a scale smaller than grid size, the
spatially averaged electric field is smaller than $\criticalEfield$.  Thus, in
general, the electrical breakdown may occur even in the region where
 $\langle \criticalEfield \rangle$ 
is smaller than  $\criticalEfield$.  Therefore, if we adopt equations 
(\ref{eqNonlinearEfield}) and (\ref{eqNonlinearMagDiff}), we may underestimate,
but not overestimate, the occurrence of electric discharges.
}

We adopted a local, Kepler-rotation shearing box that has 
radial (x), azimuthal (y) and vertical (z) axes, and solved the 
following resistive magnetohydrodynamic (MHD) equations  numerically:

\begin{eqnarray}
  \frac {\partial \rho}  {\partial t} +  \nabla \cdot \left(\rho {\mathbf v}\right) &=& 0
  \label{basicEqDensity}, \\
  \frac {\partial \mathbf v}  {\partial t} + {\mathbf v} \cdot \nabla {\mathbf v} &=&
  - \frac{1}{\rho}\nabla\left(P + \frac{B^2}{8\pi}\right) + \frac{1}{4\pi\rho}\left({\mathbf B}\cdot\nabla{\mathbf B}\right)\nonumber \\
  & & - 2{\mathbf \Omega}\times{\mathbf v}+3\Omega^2x{\hat{\mathbf x}} \label{basicEqVelocity},\\
  \frac {\partial {\mathbf B}}{\partial t} &=& - c \nabla \times {\mathbf E} \label{basicEqMagnet},
\end{eqnarray}
with isothermal EOS
\begin{eqnarray}
  P = \soundSpeed^2 \rho, \label{isothermalEqOS}.
\end{eqnarray}
Had we used the adiabatic EOS, the internal energy would have kept
growing as the linear function of time (\HGB).  Therefore, we use the
isothermal EOS to approximate the steady state attained by the cooling
processes present in the protoplanetary disks.

\begin{figure}
  \begin{center}
    \includegraphics[width=7cm]{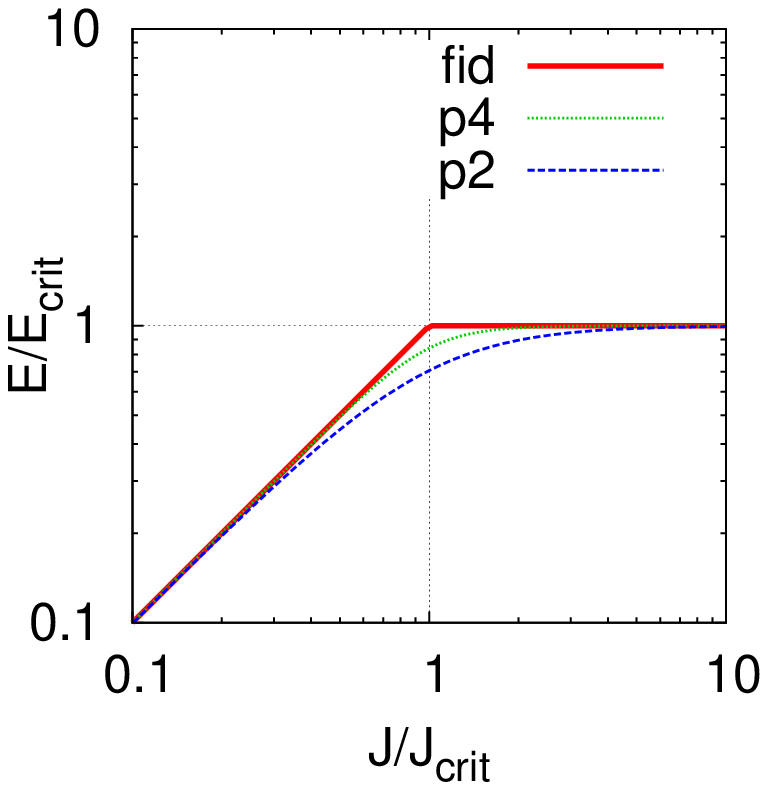}
  \end{center}
  \caption{ 
    \addspan{
      Electric field amplitude as functions of current in the 
      three different models of nonlinear Ohm's law.
      The symbols 
      {\tt fid},
      {\tt p2}, and
      {\tt p4} correspond to equation 
 (\ref{eqNonlinearMagDiff}),
 (\ref{eqNonlinearPTwo}), and
 (\ref{eqNonlinearPFour}),respectively.
    }
  }\label{figNonlinearModels}
\end{figure}

The nonlinear Ohm's law reads:
\begin{eqnarray}
  {\mathbf E} &=& -\frac{1}{c}{\mathbf v}\times{\mathbf B}+\frac{4\pi}{c^2}\magDiff\left(J\right){\mathbf J} \label{basicEqEfield},\\ \\
  {\mathbf J} &=& \frac{c}{4\pi}\nabla \times {\mathbf B} \label{basicEqCurrent},
\end{eqnarray}

\addspan{ 
  Here, we have studied three different models for nonlinear diffusivity $\magDiff(J)$.
  In addition to our fiducial model({\tt fid}),
  Eqs. (\ref{eqNonlinearEfield}) (\ref{eqNonlinearMagDiff}), we
  have studied the following two models:
  \begin{eqnarray}
    \left(\mathtt{p2}\right):\ \magDiff\left(\current\right) &=& \left(1 + \left(\frac{\criticalCurrent}{\current}\right)^{-2}\right)^{-\frac{1}{2}} \label{eqNonlinearPTwo}, \\
    \left(\mathtt{p4}\right):\ \magDiff\left(\current\right) &=& \left(1 + \left(\frac{\criticalCurrent}{\current}\right)^{-4}\right)^{-\frac{1}{4}} \label{eqNonlinearPFour}.
  \end{eqnarray}
  The electric field as functions of current density in these three models are shown in
  Figure \ref{figNonlinearModels}.
}

Following \HGB, we set up our numerical initial conditions as
follows.
{
 We use the disk scaleheight $H$ as the unit length.
}
The box size is $(L_x, L_y, L_z) = (1, 2 \pi, 1)$.  First,
we set the average values $\density_0 = 1$ and $\pressure_0 =
10^{-6}$ to every mesh, and let fluid velocity to be at rest in
shearing box frame; \addspan{$(v_x, v_y, v_z = 0,-(3/2)\Omega x,0)$}. Here,
${\soundSpeed} = 10^{-3}$, and also $\Omega = 10^{-3}$. 

Next, we introduce random perturbations in density, pressure, and velocity.
The density and pressure perturbations are in proportion so that the isothermal EOS 
is met, and the amplitude is $\delta\rho/\rho_0 = \delta P/P_0 = 2.5\times10^{-2}$.
We perturb the velocity component-wise, with the amplitude 
${\delta v_i} = 5\times 10^{-3}\soundSpeed$ for each.

We use the following units of magnetic field and electric current:
\begin{eqnarray}
  \equipartitionMagnet &\equiv& \sqrt{8 \pi P_0} \label{eqUnitMagnet} ,\\
  \equipartitionCurrent &\equiv& \frac{{c \equipartitionMagnet} }{ 4\pi H} \label{eqUnitCurrent} ,
\end{eqnarray}
and the scale height
\begin{eqnarray}
  \scaleHeight &=& \frac{\soundSpeed}{\orbitalOmega} \label{eqScaleHeight} .
\end{eqnarray}

We set uniform magnetic field in the $z$-direction, and express the initial field strength by 
the plasma beta, 
\begin{eqnarray}
  \beta \equiv {\equipartitionMagnet}^2/{B_{z0}}^2  = 8\pi P_0/ {B_{z0}}^2 \label{eqPlasmaBeta} .
\end{eqnarray}

The plasma beta satisfy the following relation between the sound speed $\soundSpeed$ and the Alfv\'en velocity
along the magnetic field $\alfvenVelocityZ$:
\begin{eqnarray}
  \beta = \frac {2 \soundSpeed^2}{{\alfvenVelocityZ}^2}  \label{eqPlasmaBeta2} .
\end{eqnarray}

{
We define the magnetic Reynolds number as:
\begin{eqnarray}
\magneticReynolds \equiv {\alfvenVelocityZ}^2 / \magDiff_0 \Omega,
\end{eqnarray}
using the Alfv\'en velocity $\alfvenVelocityZ = B_{z0}/\sqrt{4\pi \density}$
set by the initial vertical magnetic field. 
This is in accordance with \SMUN\ and \ISSS\, while some literature adopts
different definition
\citep[e.g.  $\magneticReynolds \equiv {\soundSpeed}^2 / \magDiff_0 \Omega $ in ][]{fleming_effect_2000}
}.

We have used Athena \citep{gardiner_unsplit_2005,gardiner_unsplit_2008}, an
open-source MHD code for our simulations.

\subsection{Simulations Procedure}\label{sectionSimulations}

We vary the initial magnetic field strength and the diffusivity
models, and we classify each set of parameters as either $\bigcirc${\tt active
  zone}, $\times${\tt dead zone}, or $\bigtriangleup${\tt sustained zone}. The experiment
method and the definition of the three classes are given in this
section.

The parameters we have investigated are the initial vertical field strength
(represented by plasma $\beta$), the linear diffusivity (represented by
magnetic Reynolds number $\magneticReynolds$), and the critical current
$\criticalCurrent$.  The range of the survey was $400 \leq \beta \leq
25600$, $0.002 \leq \magneticReynolds \leq 2$ and $0.01 \leq
\criticalCurrent / \equipartitionCurrent \leq 100$. \addspan{In addition, the limiting cases of
$\magneticReynolds = \infty$ and
$\criticalCurrent / \equipartitionCurrent = \infty$, that respectively correspond to
ideal MHD models and linear Ohm's law models,
 are studied for comparison with the literature.
}

For each value of $\beta$, we prepared the initial condition as described in
section \ref{sectionNumericalSetup}, and continued the simulation for 10
orbits ($t = 20\pi/\Omega$), at first with magnetic diffusivity turned
off ($\magDiff(J) = 0$).  While running the simulations, we
created restart data for every periodic points ($t = 2n\pi/\Omega$
where $n$ is an integer). The condition allowed the MRI to grow and
saturate in about 5 orbits ($t = 10\pi/\Omega$).

Then, for each pair of ($\magneticReynolds$, $\criticalCurrent /
\equipartitionCurrent$), we turned on the diffusivity and re-started
the simulation either from the initial laminar flow ($t = 0$) or the
saturated MRI states at 8, 9 and 10 orbit ($t = 16\pi/\Omega,
18\pi/\Omega, 20\pi/\Omega$). We numerically evolved them until they
reach 20 orbit ($t = 40\pi/\Omega$).
The reason why we have adopted three different MRI saturated initial conditions
($t = 16\pi/\Omega, 18\pi/\Omega, 20\pi/\Omega$) for each set of the
parameters is that a `turbulent initial condition' is not unique; therefore
we need to test if our results depend on the choice of the initial condition or not.

During each simulation run, we recorded the space averages of
physical quantities as the functions of time, such as magnetic energy
density $B^2$, the Reynolds and Maxwell stress $\rho v_x \delta v_y$,
${ -B_x B_y}/{4\pi}$, and the squared current $J^2$. After the simulations we
studied the time average of the quantities.  For a physical quantity
$A$, we denote its space and time average by $\langle A \rangle$ and $
\overline{A}$, respectively. Their definitions are as follows:
\begin{eqnarray}
  \langle A \rangle &\equiv&
  \frac{\int\mathit{dx}\int\mathit{dy}\int\mathit{dz}A} 
       {\int\mathit{dx}\int\mathit{dy}\int\mathit{dz}}, \\
  \overline{A} &\equiv&
  \frac{\int \mathit{dt} A}
       {\int \mathit{dt}}.
\end{eqnarray}
The space average is taken for the entire computational domain ($0 < x
< L_x$, $0 < y < L_y$, $0 < z < L_z$),
\addspan{ numerical resolution is 
$(N_x,N_y,N_z) = (64, 128, 64)$,}
 and the time average is taken
for the last five orbit ($30\pi < t < 40\pi $) unless otherwise
mentioned.  The statistics for the important sets of parameters are
presented at the end of the paper.

\begin{table}
  \begin{tabular}{c|cc}
    zone & from laminar? & from ideal MRI? \\
    \hline
    $\bigcirc$active& unstable & unstable \\
    $\times$dead& stable & stable \\
    $\bigtriangleup$sustained& stable & unstable
  \end{tabular}
  \caption {
    The zone names and the meaning of the symbols in Figures \ref{figSMRIsurvey}
    and \ref{figSMRIsurvey2}.
  }\label{tblSymbol}
\end{table}

\begin{figure*}
  \begin{center}
    \begin{tabular}{lcc}
      & restart from $t = 0$
      & restart from $t = 16\pi/\Omega$\\
      \parbox{0.4cm}{(a) \\ $\bigcirc$} &
      \raisebox{-4.02cm}{\includegraphics[width=7cm]{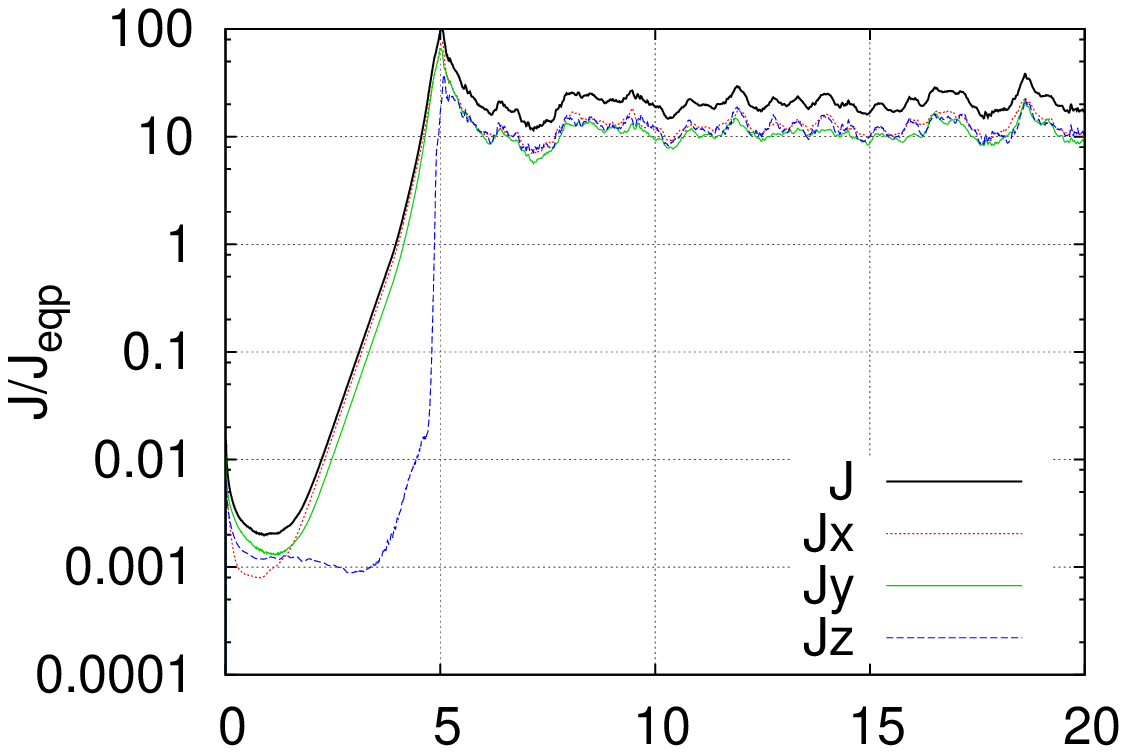}} & 
      \raisebox{-4.02cm}{\includegraphics[width=7cm]{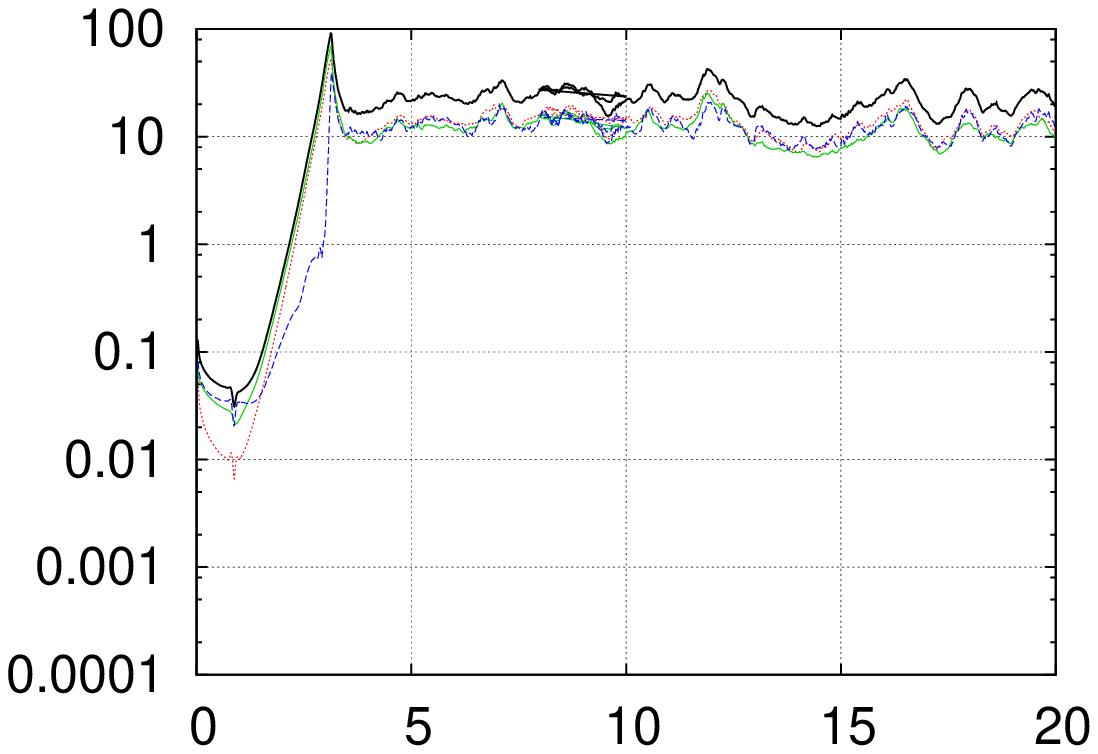}} \\
      \parbox{0.4cm}{(b) \\ $\bigtriangleup$} &
      \raisebox{-4.02cm}{\includegraphics[width=7cm]{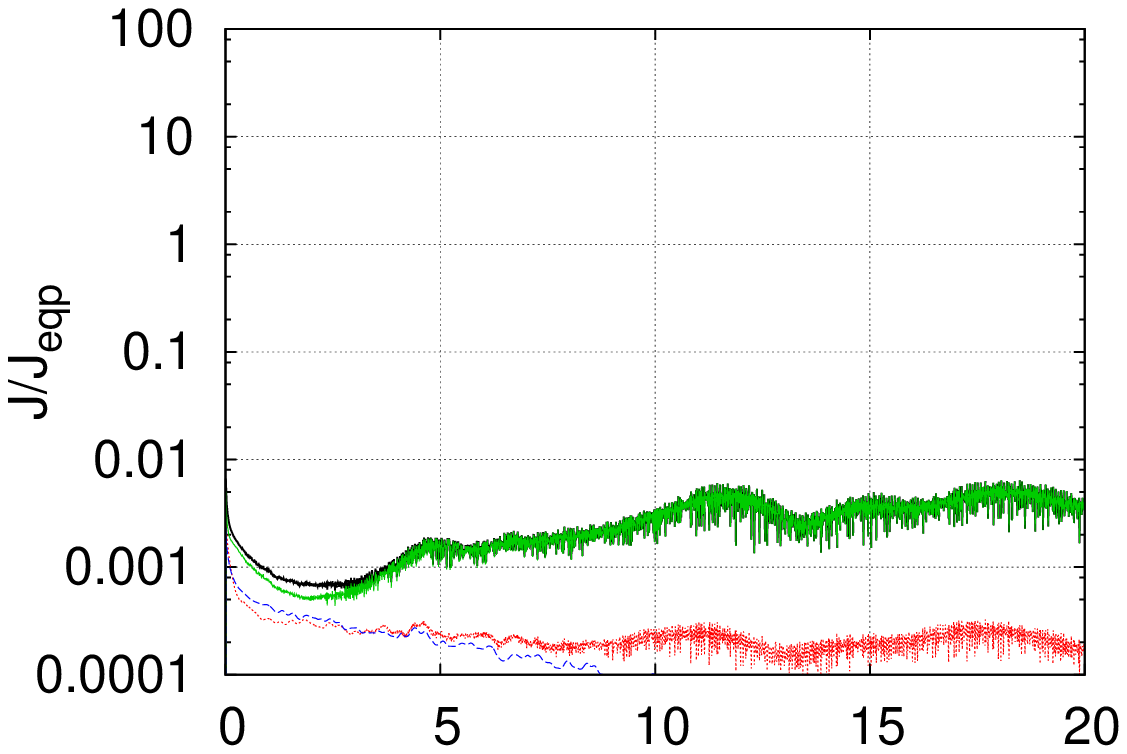}} & 
      \raisebox{-4.02cm}{\includegraphics[width=7cm]{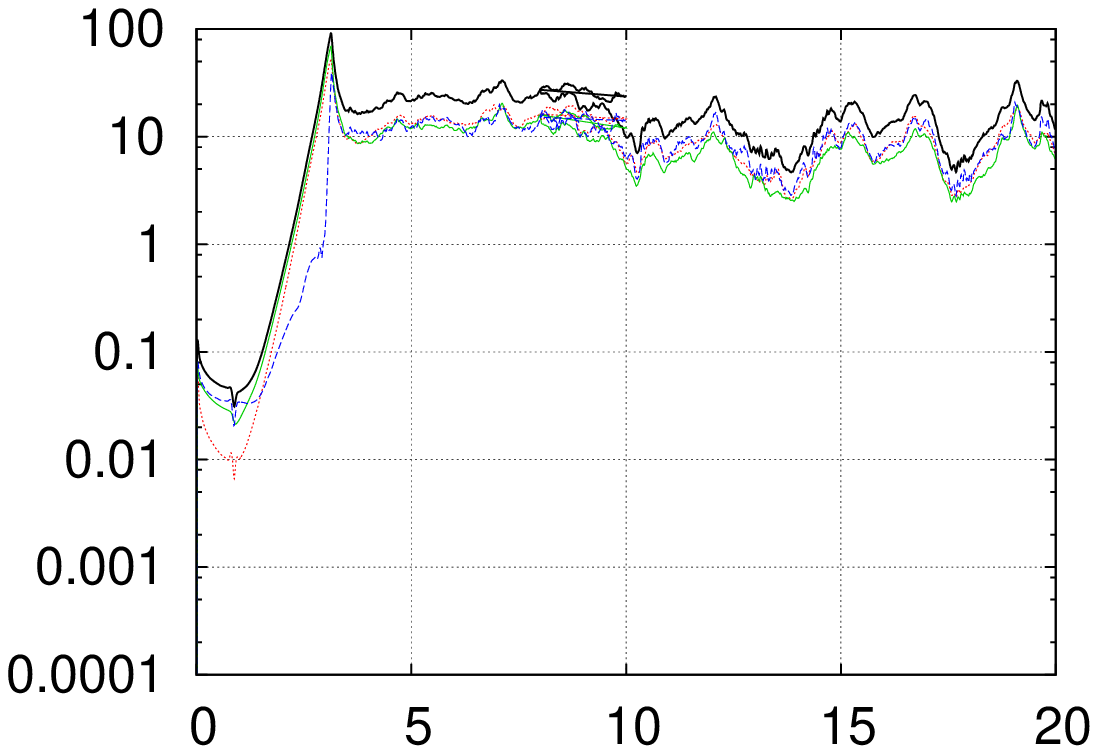}} \\
      \parbox{0.4cm}{(c) \\ $\times$} &
      \raisebox{-4.02cm}{\includegraphics[width=7cm]{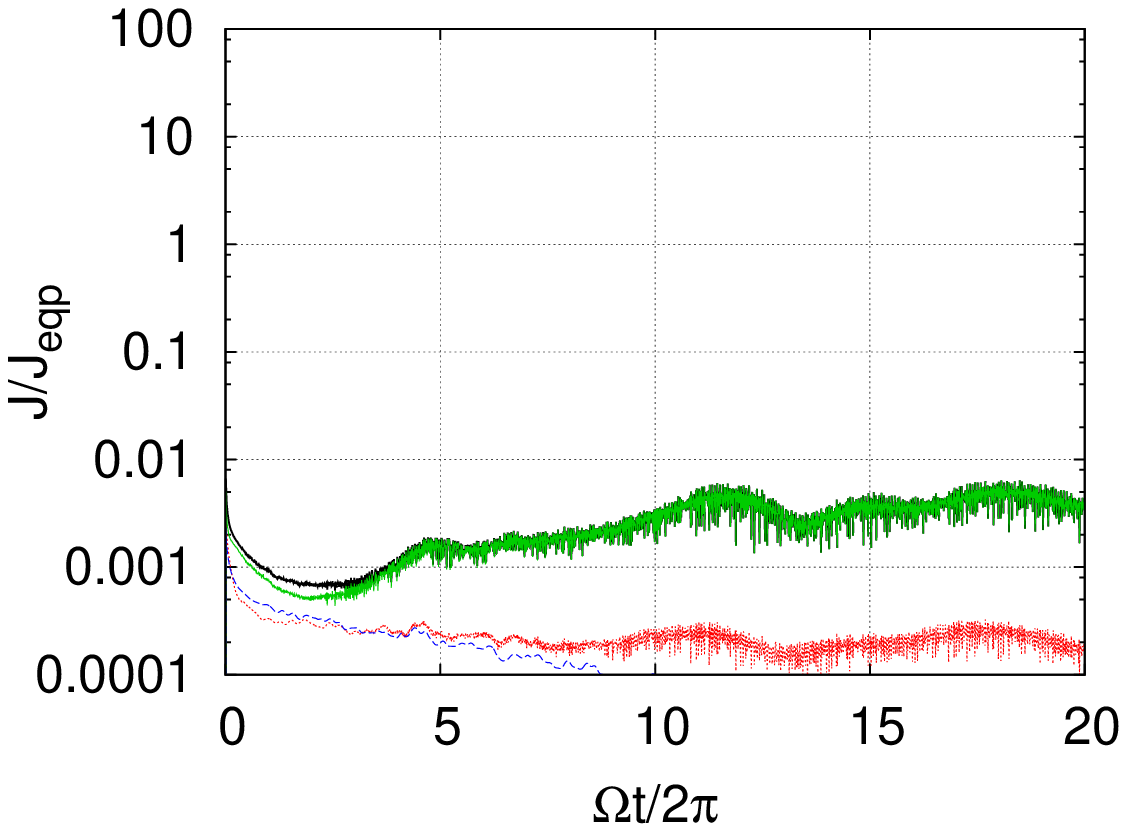}} & 
      \raisebox{-4.02cm}{\includegraphics[width=7cm]{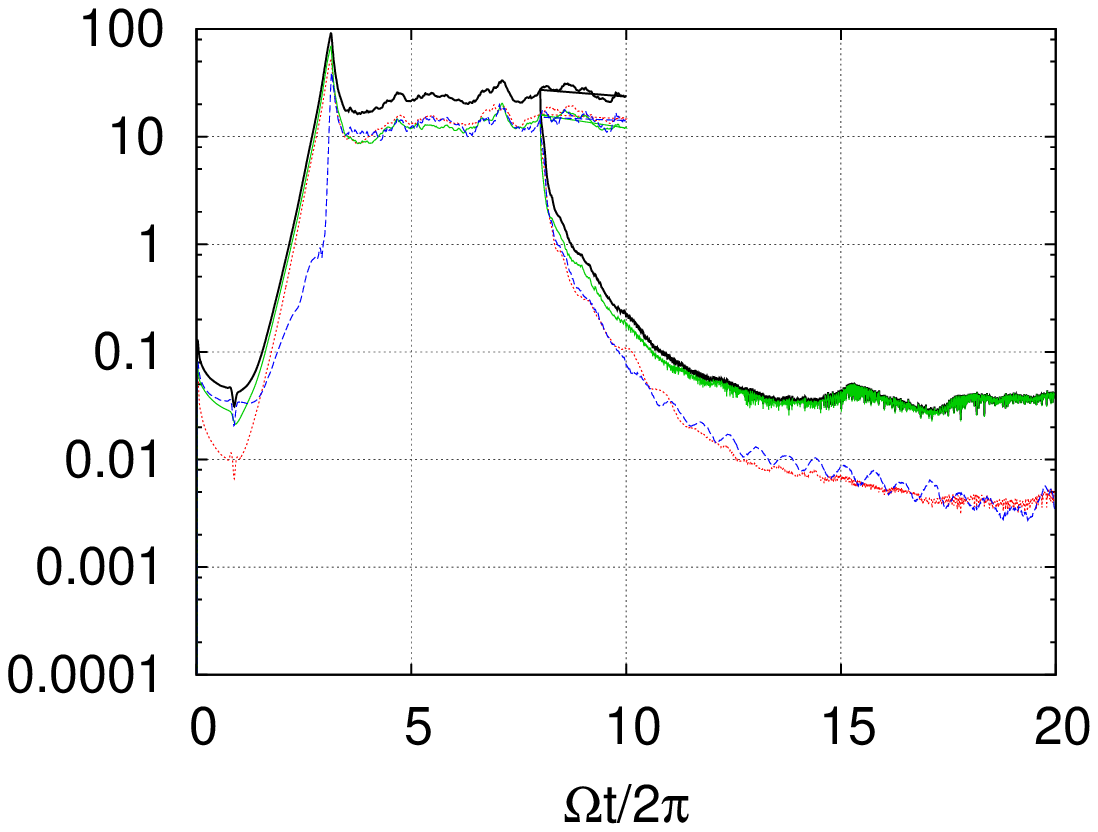}} \\
    \end{tabular}
  \end{center}
  \caption {
    Time evolution of the averaged current density for three magnetic diffusivity models.
    Models are 
    (a) $\beta=400$, $R_M=0.6$, $\criticalCurrent/\equipartitionCurrent = 1$ 
    (b) $\beta=400$, $R_M=0.2$, $\criticalCurrent/\equipartitionCurrent = 1$  
    (c) $\beta=400$, $R_M=0.2$, $\criticalCurrent/\equipartitionCurrent = 10$ .
    The graphs show typical current behavior for
    (a) $\bigcirc$       active zone,
    (b) $\bigtriangleup$ sustained zone, or
    (c) $\times$         dead zone.
  }\label{figSMRIABC}
\end{figure*}

Using the average values, we classify each set of the parameter
($\beta$, $\magneticReynolds$, $\criticalCurrent /
\equipartitionCurrent$) as follows (c.f. Table
\ref{tblSymbol}). First, a parameter is in $\bigcirc${\tt active zone}
if the MRI is observed both in the simulation started from
laminar flow as well as in all of the three simulations started from
the saturated MRI states. Second, a parameter is in $\times${\tt dead
  zone} if the instability is not observed neither in the
simulation started from laminar flow as well as in any of the three
simulations started from the saturated MRI states. Finally, a parameter
is in $\bigtriangleup${\tt sustained zone}, if the MRI {\em is} observed at
in all the three simulations started from the saturated MRI
states, but {\em not} at the end of the simulation started from laminar
flow.
\subsection{The Result of Shearing-Box Simulations}

{
The typical behavior of the current for the active zone, 
dead zone and sustained zone are in Figure \ref{figSMRIABC}.
From the simulations we have observed that the three classes (active, dead and sustained)
are exhaustive:
that the runs started from 8,9 and 10 orbit always agree in terms of the classification;
and that if the MRI dies when starting from saturated initial condition, it also does not activate starting from
laminar initial condition.
}

To classify the active, dead and sustained zone,
we need to assess the magnetorotational instability of the system, so we introduce
the following criteria for quantitative assessment.
We say that the system is magnetorotationally unstable if the
averaged current $\langle \current^2 \rangle^{1/2}$ is greater than $0.1
\equipartitionCurrent$, and stable if otherwise.
Here, the time average is taken for the last five orbit ($30\pi < t < 40\pi $).
 We have learned from the
simulations that the quantity $\langle \current^2 \rangle^{1/2}$ is a good
indicator for the stability, since it either fluctuates around mean
value $\langle \current^2 \rangle^{1/2} \simeq 10
\equipartitionCurrent$ (unstable) or go under $\langle \current^2
\rangle^{1/2} < 0.1 \equipartitionCurrent$ almost monotonically with
little vibration (stable), and there is no ambiguity between the two;
c.f.  Figure \ref{figSMRIABC}.  

However, as we make $\magneticReynolds$ smaller, 
the diffusion timescale becomes shorter, and the wall clock time for
the simulations until $t = 40\pi$ becomes impractically larger.
Therefore, for the parameter range $\magneticReynolds < 0.01$,
we terminate the simulations at $1.5\times 10^6$ cycles
and determine the class by extrapolations.

\begin{figure*}
  \begin{center}
    \includegraphics[width=7cm]{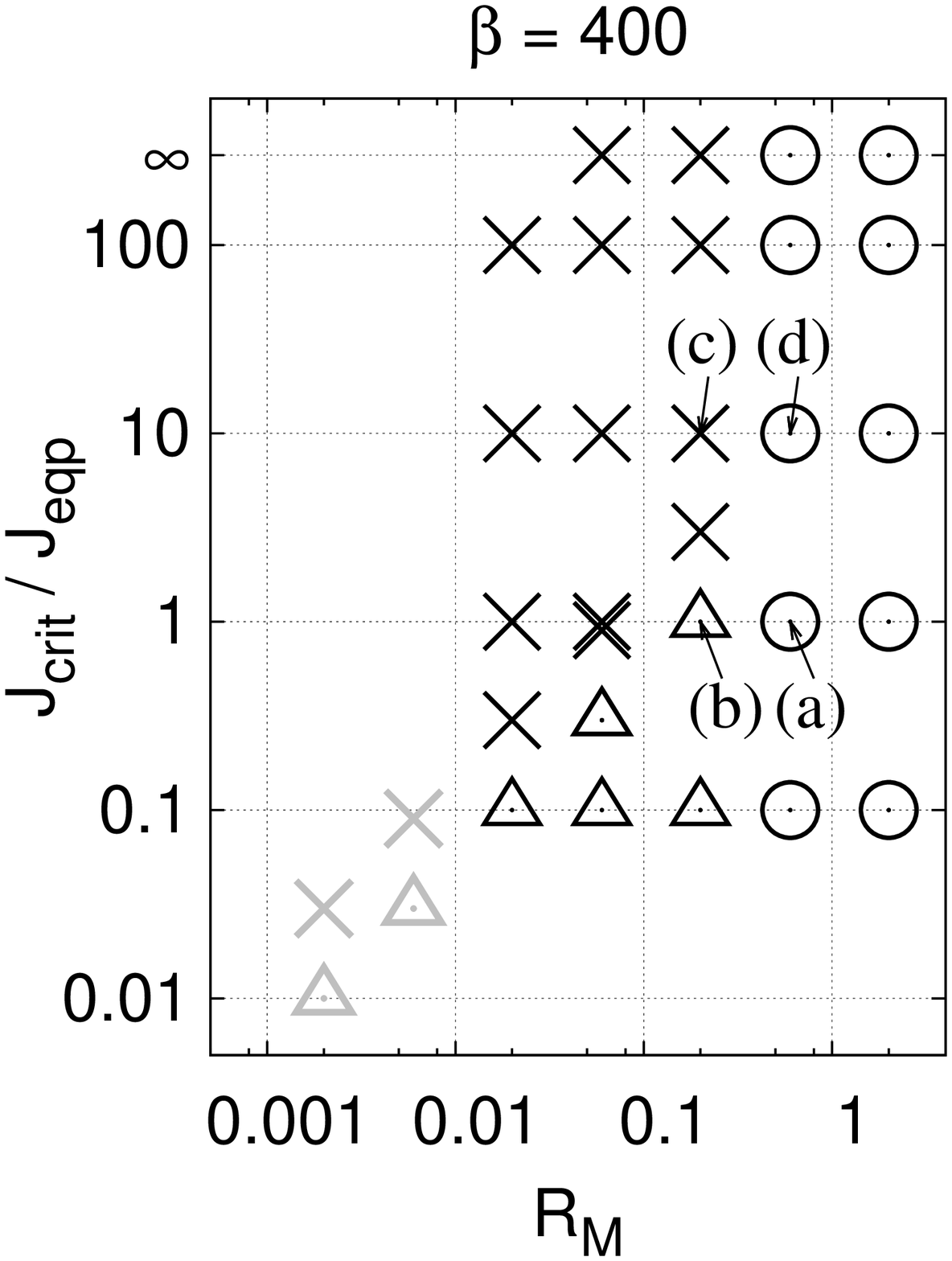}
    \includegraphics[width=7cm]{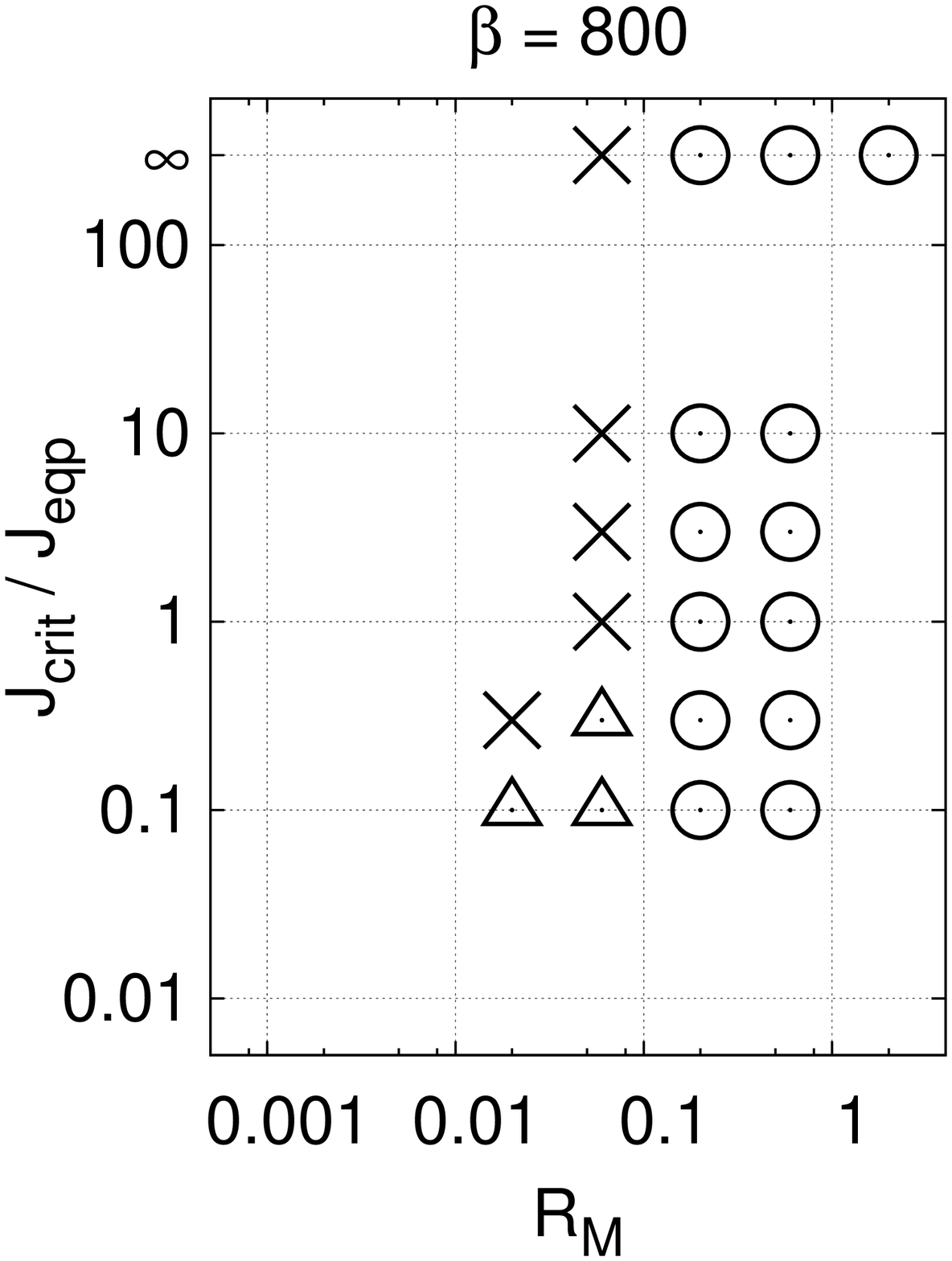}
    \includegraphics[width=7cm]{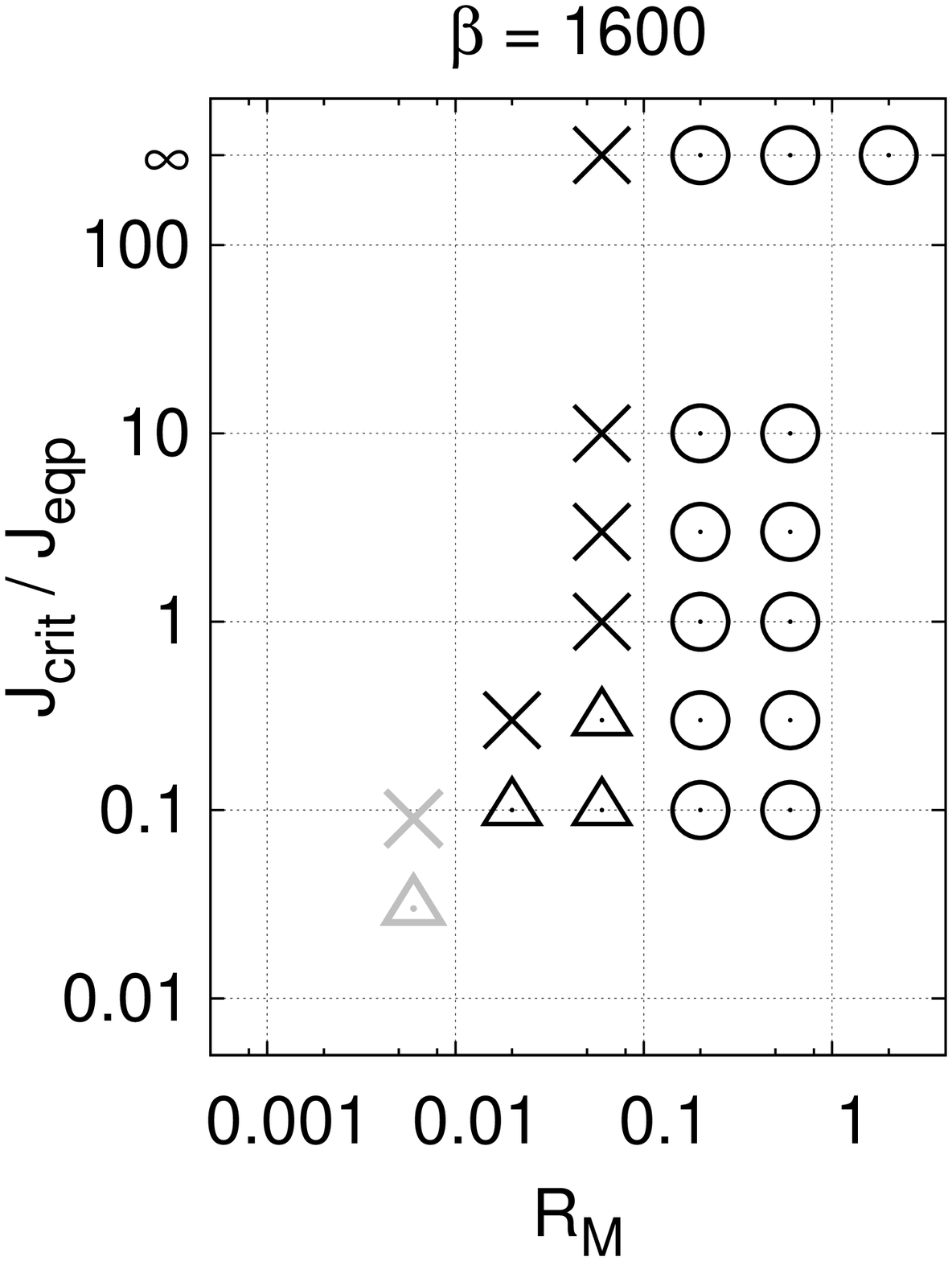}
    \includegraphics[width=7cm]{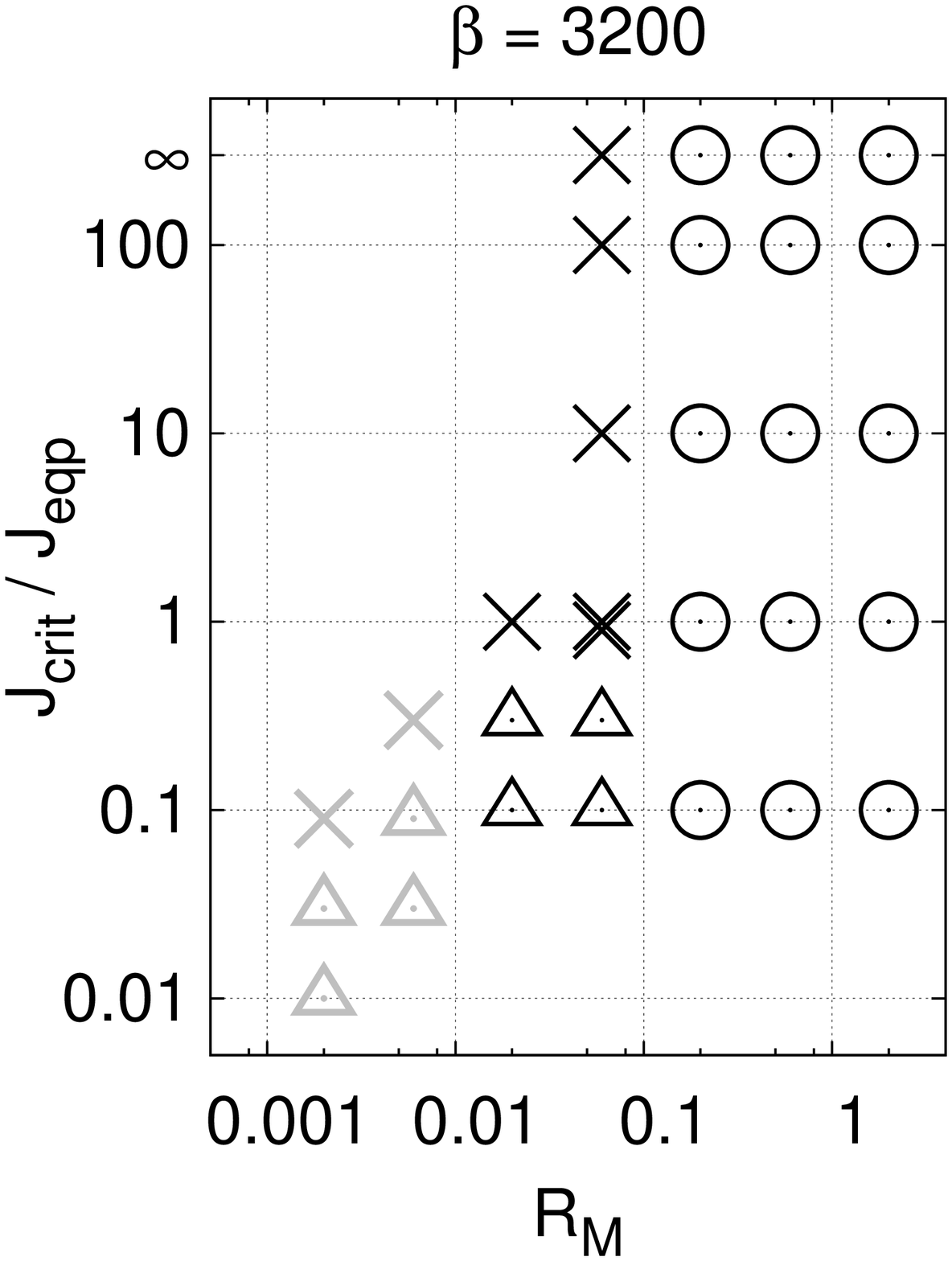}
  \end{center}
  \caption { Distribution of the sets of parameters ($\beta$,
    $\magneticReynolds$, $\criticalCurrent / \equipartitionCurrent$)
    in our simulations. We classify each of them as either $\bigcirc$
    active zone, $\bigtriangleup$ sustained zone, or $\times$ dead
    zone.  This page includes the data for $400 \leq \beta \leq 3200$.
    The parameters classified by extrapolations are marked by light
    gray symbols.  }\label{figSMRIsurvey}
\end{figure*}

\begin{figure*}
  \begin{center}
    \includegraphics[width=7cm]{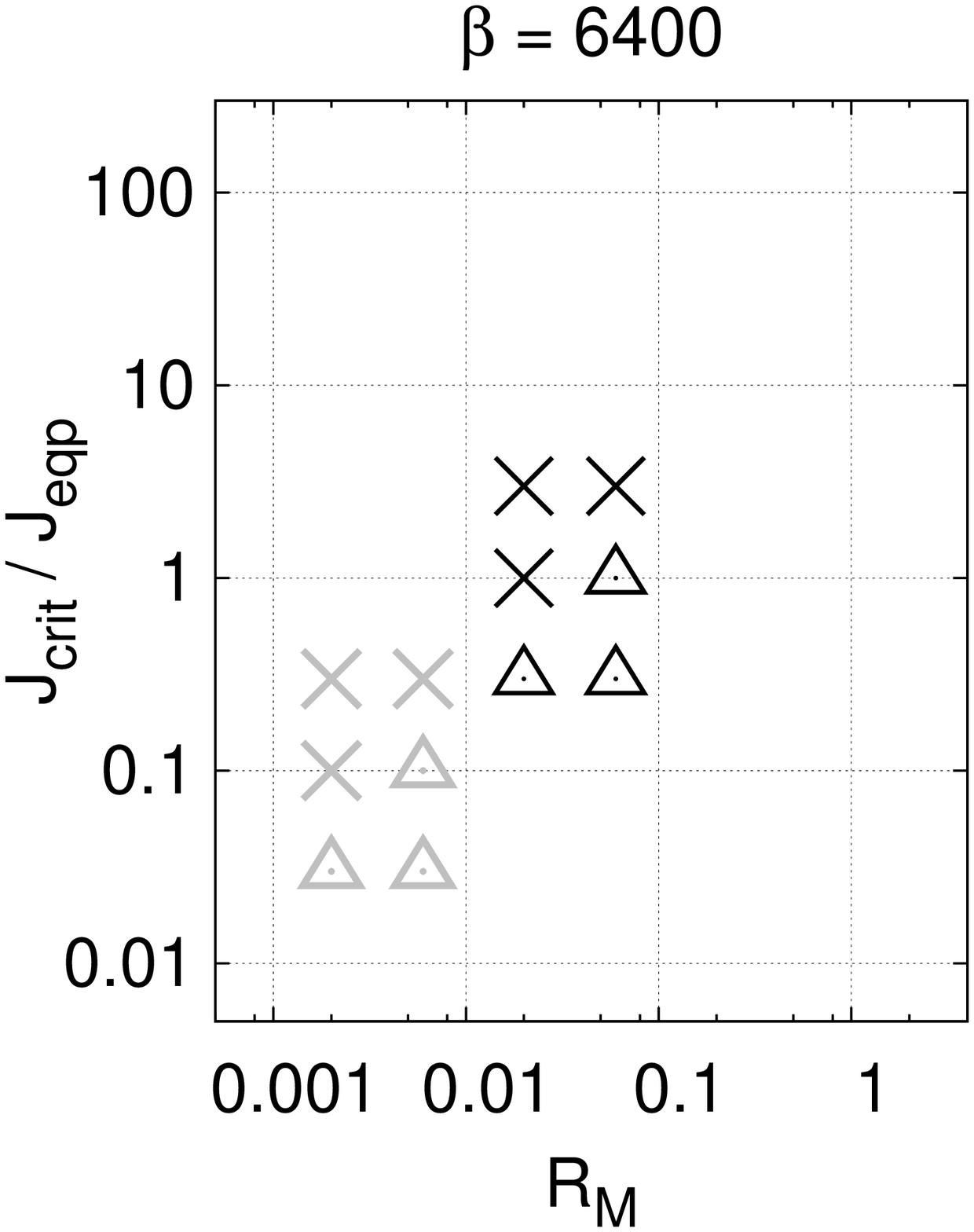}
    \includegraphics[width=7cm]{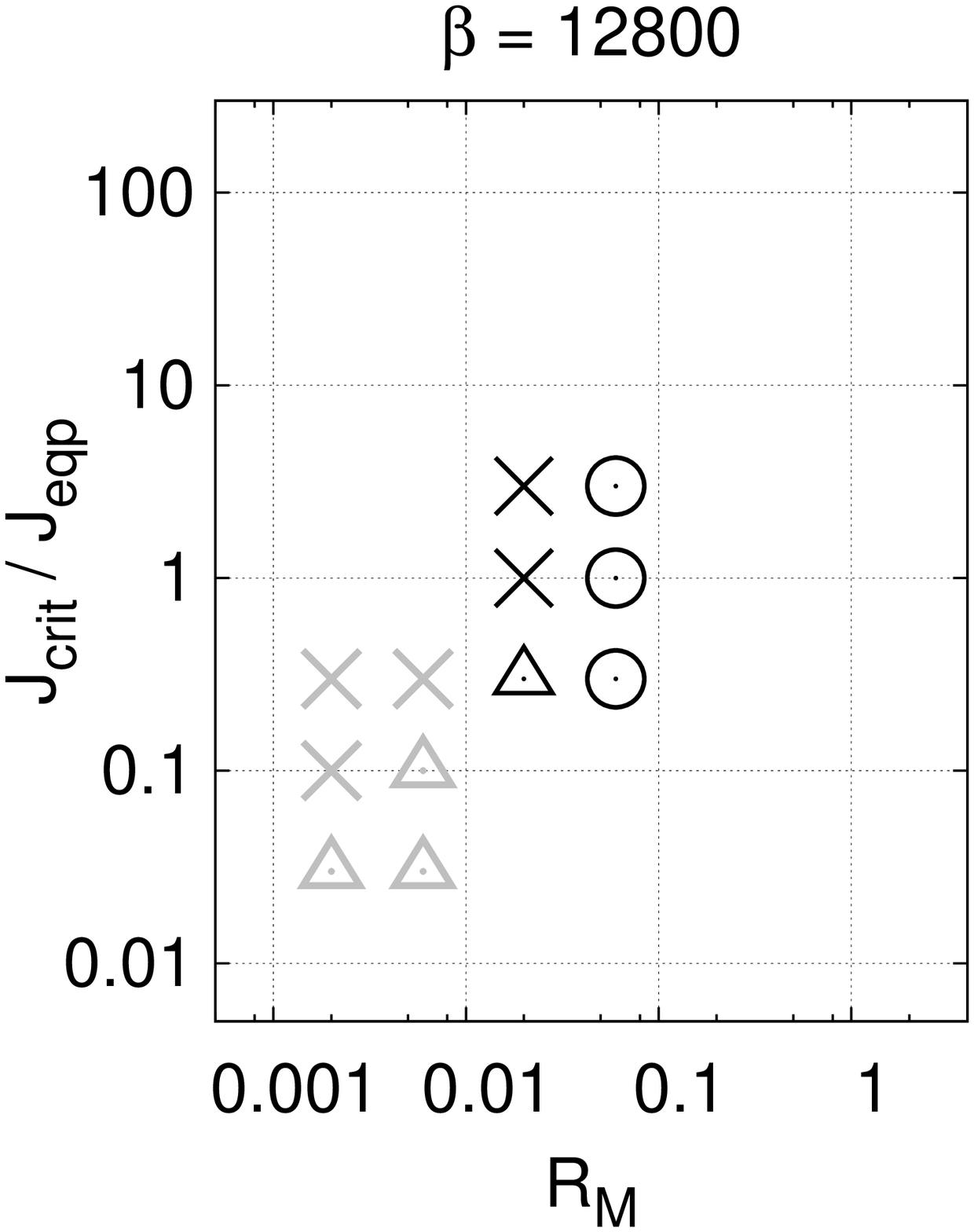}
    \includegraphics[width=7cm]{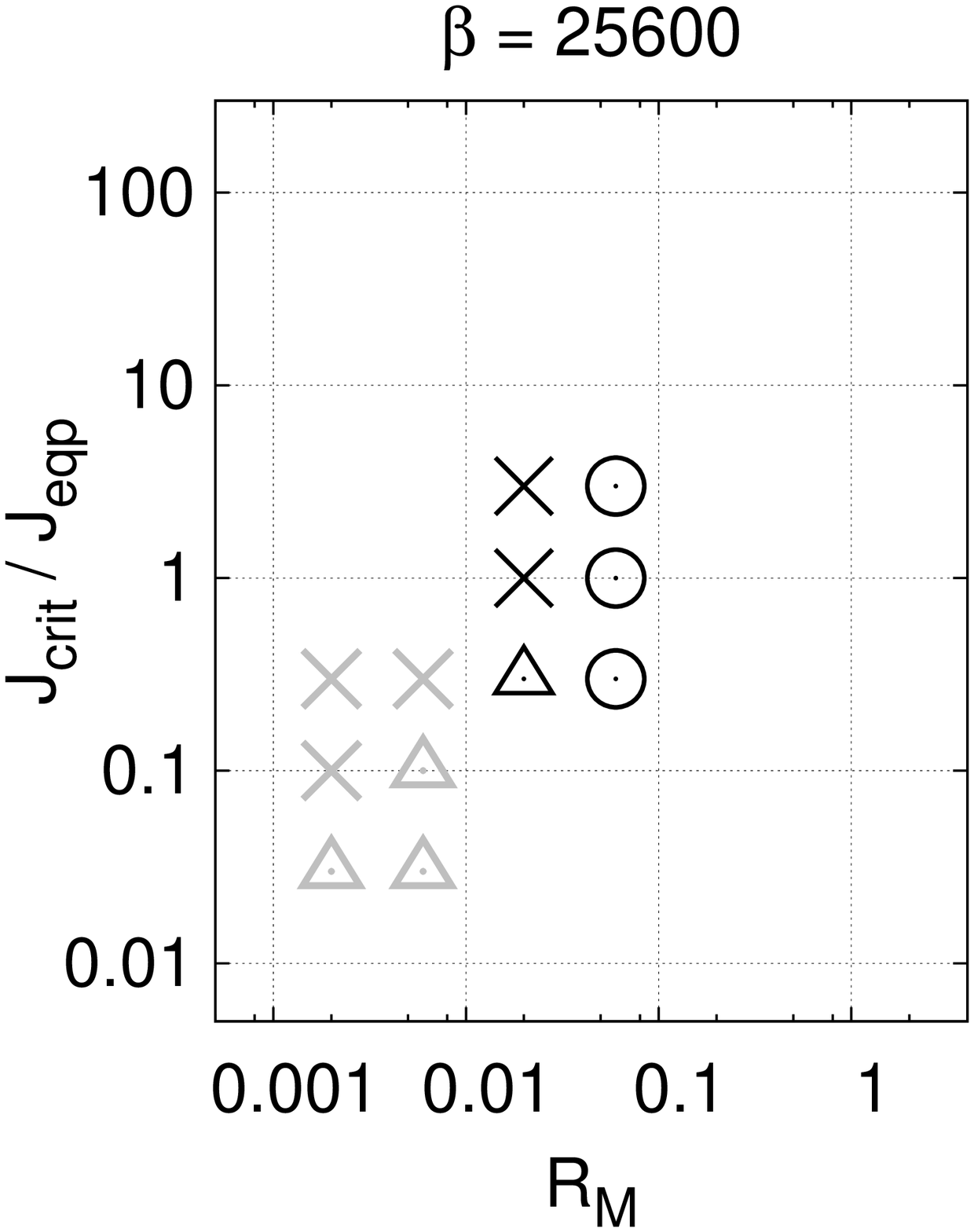}
  \caption { 
    Continued from Figure \ref{figSMRIsurvey}, the data for $6400 \leq \beta \leq 25600$.
  } \label{figSMRIsurvey2}
  \end{center}
\end{figure*}

Figures \ref{figSMRIsurvey} and  \ref{figSMRIsurvey2} show the result of our parameter survey.  We
found that sustained zone exist --- the MRI does exhibit hysteresis
behavior for a certain set of parameters. 

\begin{figure*}
  \begin{center}
    \begin{tabular}{lcc}
      & restart from $t = 0$
      & restart from $t = 16\pi/\Omega$\\
      \parbox{0.4cm}{(a) \\ $\bigcirc$} &
      \raisebox{-4.02cm}{\includegraphics[width=7cm]{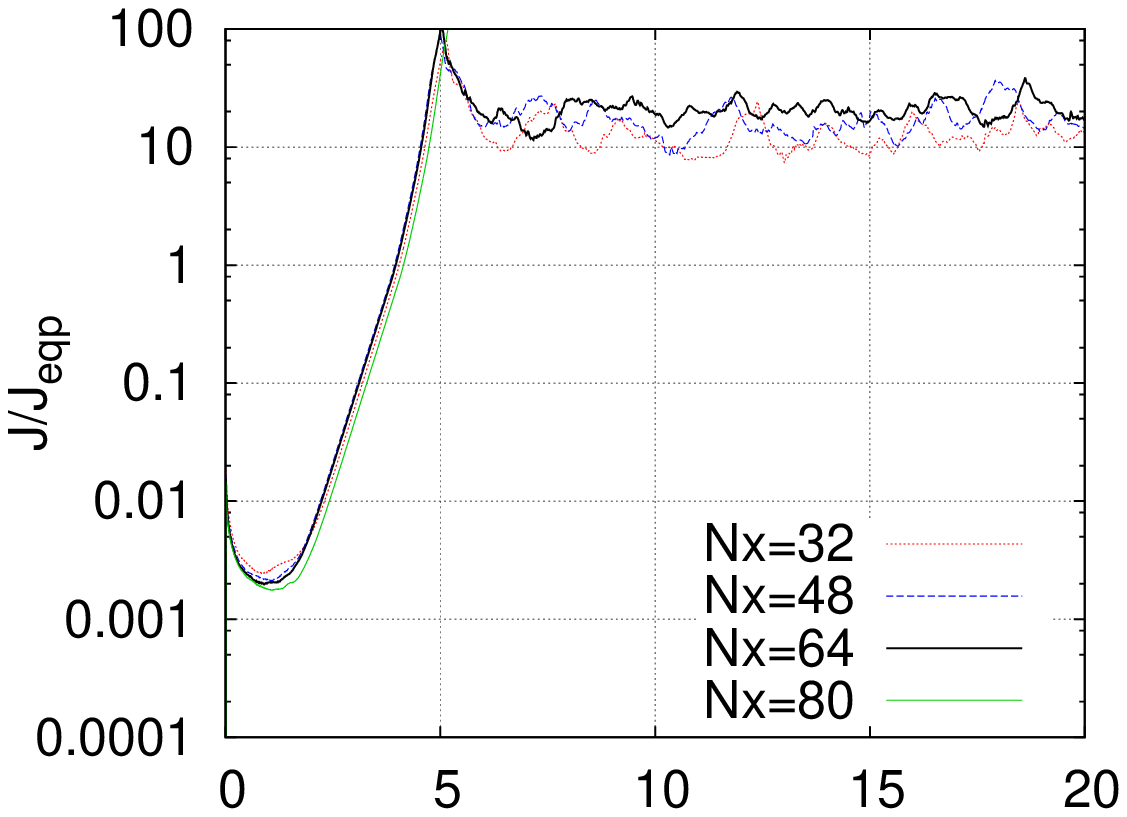}} & 
      \raisebox{-4.02cm}{\includegraphics[width=7cm]{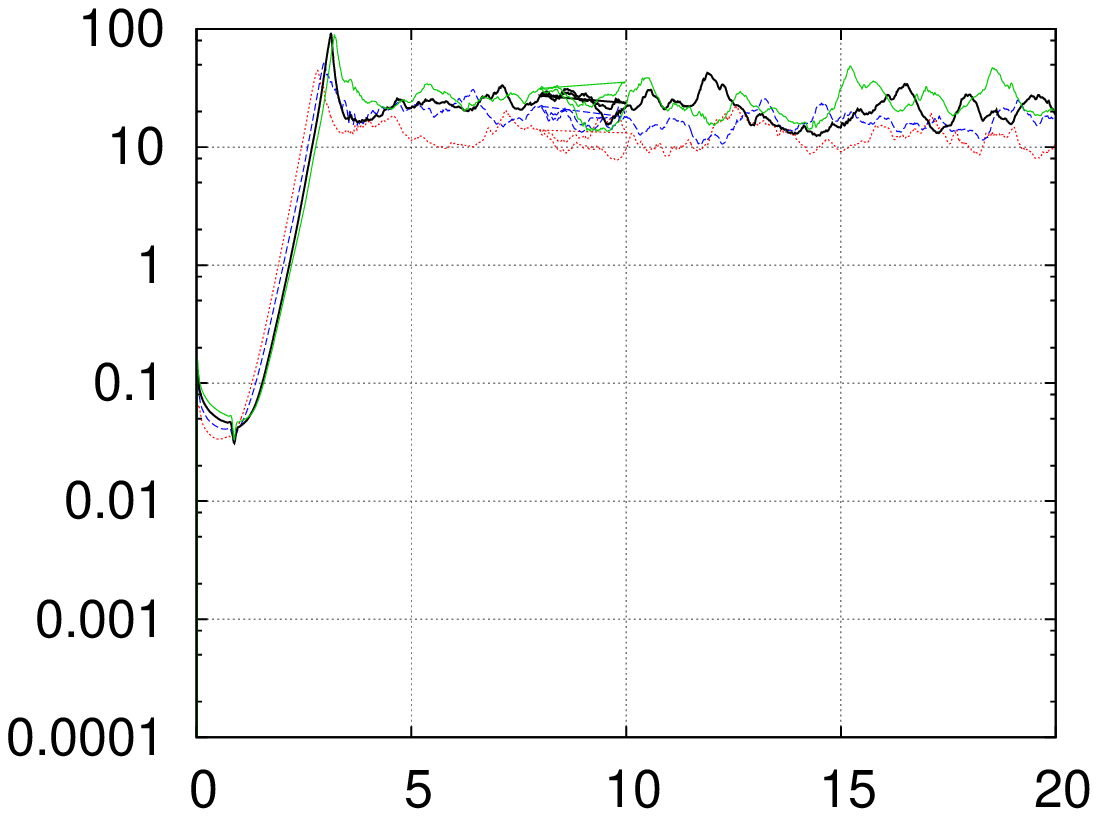}} \\
      \parbox{0.4cm}{(b) \\ $\bigtriangleup$} &
      \raisebox{-4.02cm}{\includegraphics[width=7cm]{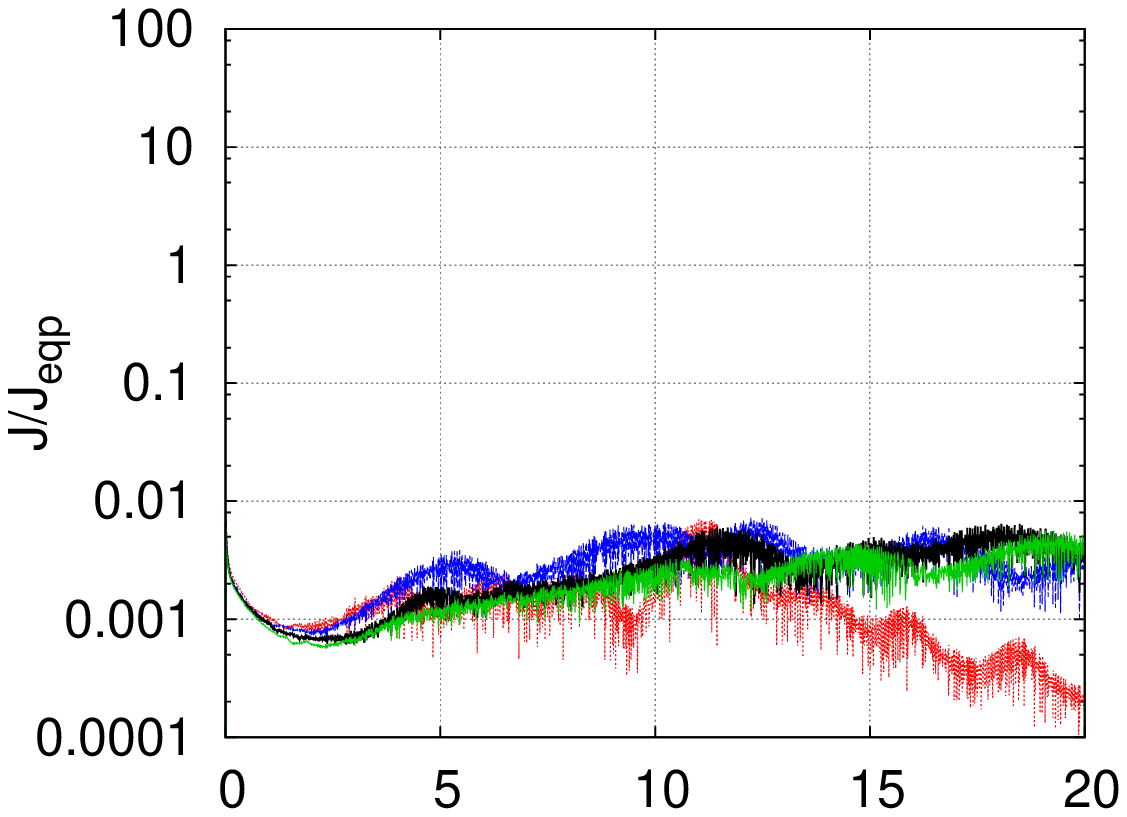}} & 
      \raisebox{-4.02cm}{\includegraphics[width=7cm]{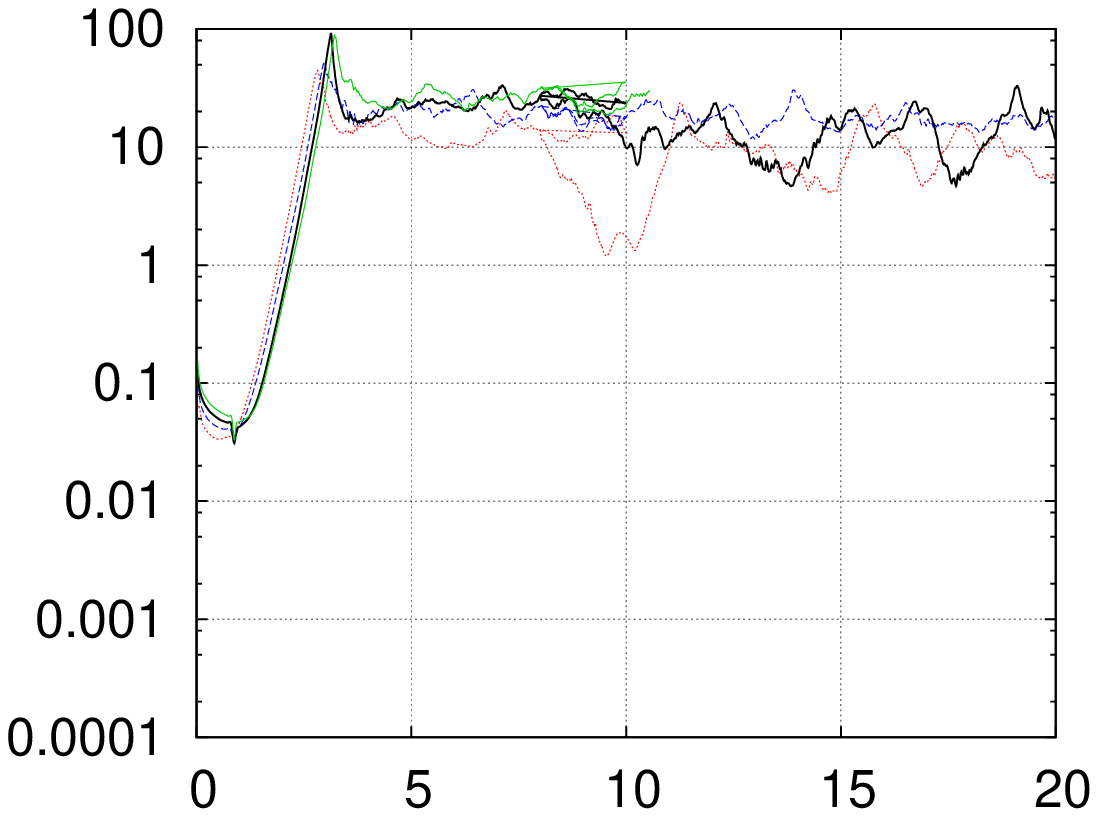}} \\
      \parbox{0.4cm}{(c) \\ $\times$} &
      \raisebox{-4.02cm}{\includegraphics[width=7cm]{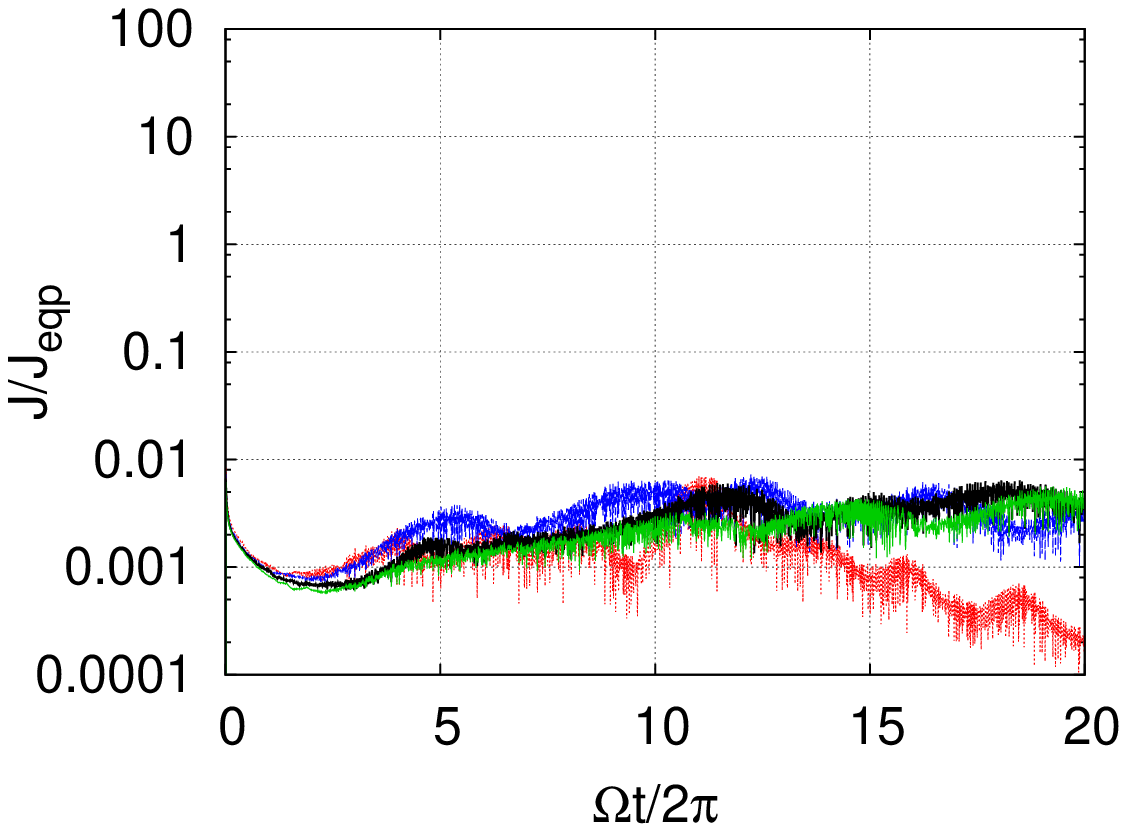}} & 
      \raisebox{-4.02cm}{\includegraphics[width=7cm]{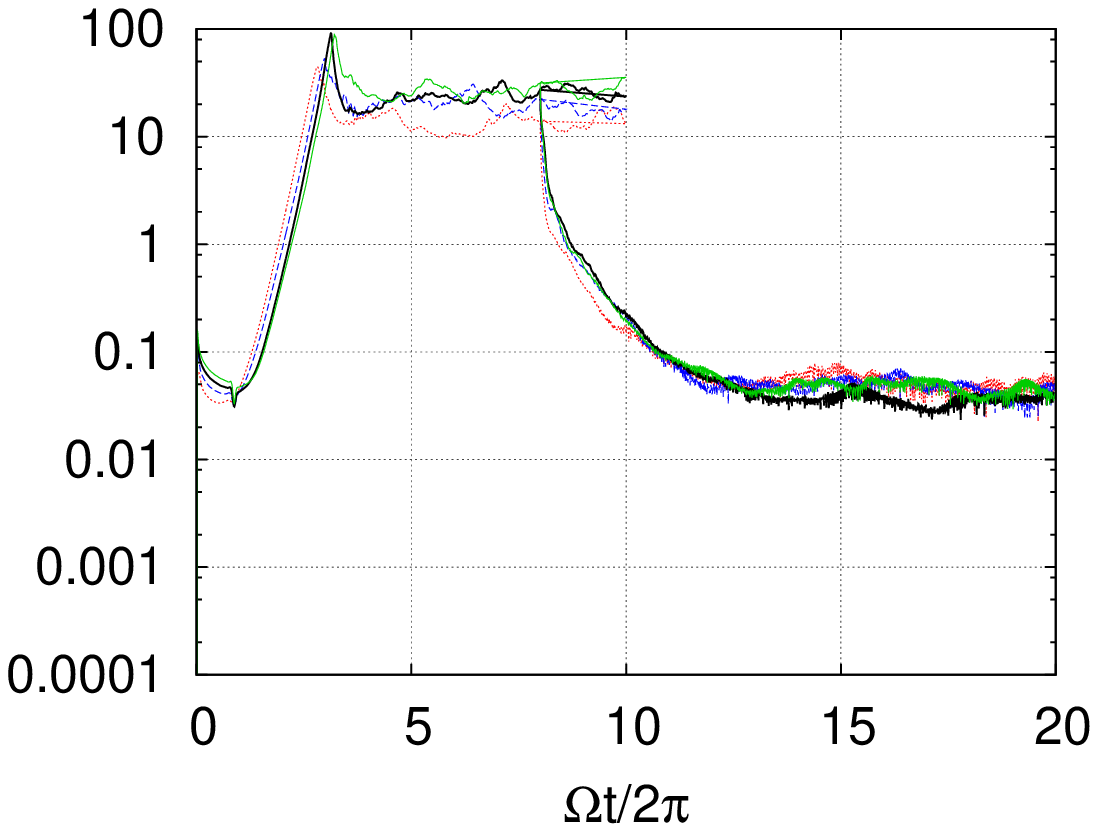}} \\
    \end{tabular}
  \end{center}
  \caption {
    \addspan {
      Dependence of the MRI behavior on the resolution.
      The development of electric current over time in the 
    simulations using the different numerical resolution
    $N_x = 32,48,64$(fiducial) and $80$ while keeping the aspect ratio $N_x : N_y : N_z = 1:2:1$, 
    for three different set of parameters
    (a) $\beta=400$, $R_M=0.6$, $\criticalCurrent/\equipartitionCurrent = 1$ ,
    (b) $\beta=400$, $R_M=0.2$, $\criticalCurrent/\equipartitionCurrent = 1$ and 
    (c) $\beta=400$, $R_M=0.2$, $\criticalCurrent/\equipartitionCurrent = 10$ .
    They are typical parameter sets for 
    (a) $\bigcirc$       active zone,
    (b) $\bigtriangleup$ sustained zone, and
    (c) $\times$         dead zone, respectively.
    }
  }\label{figConvergence}
\end{figure*}

\begin{figure*}
  \begin{center}
    \begin{tabular}{lcc}
      & restart from $t = 0$
      & restart from $t = 16\pi/\Omega$\\
      \parbox{0.4cm}{(a) \\ $\bigcirc$} &
      \raisebox{-4.02cm}{\includegraphics[width=7cm]{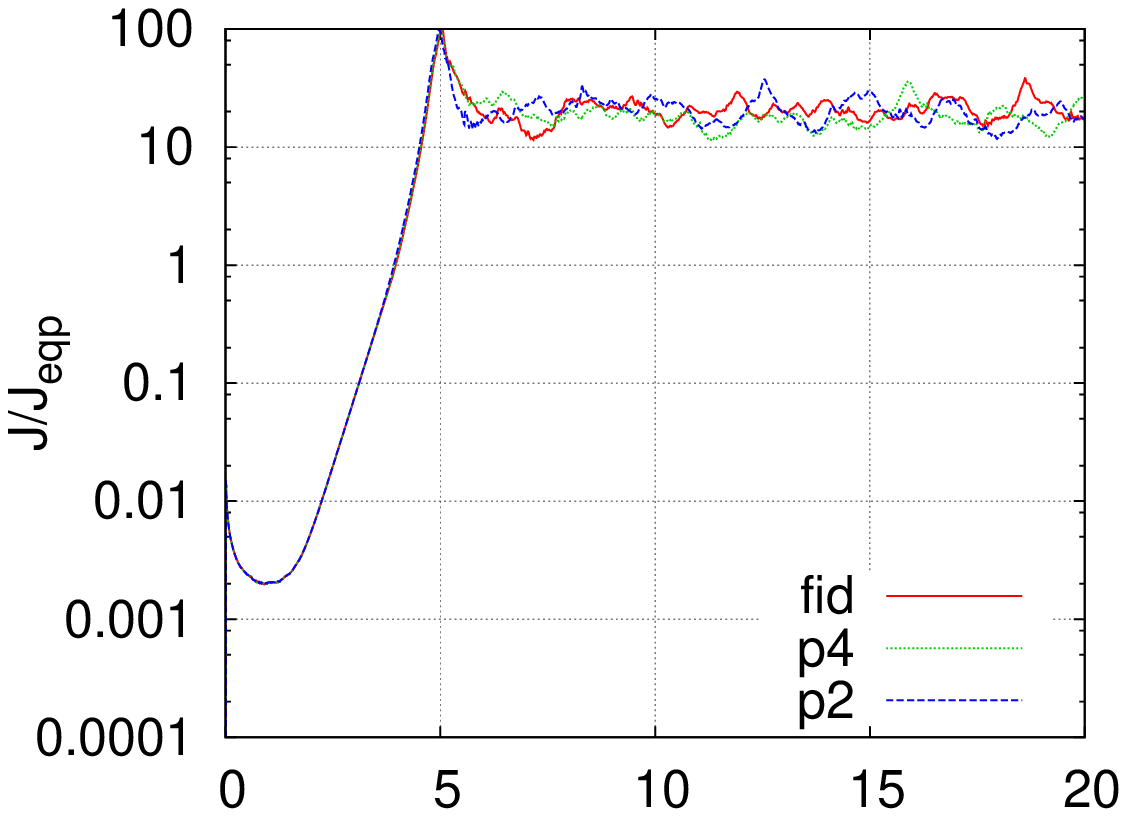}} & 
      \raisebox{-4.02cm}{\includegraphics[width=7cm]{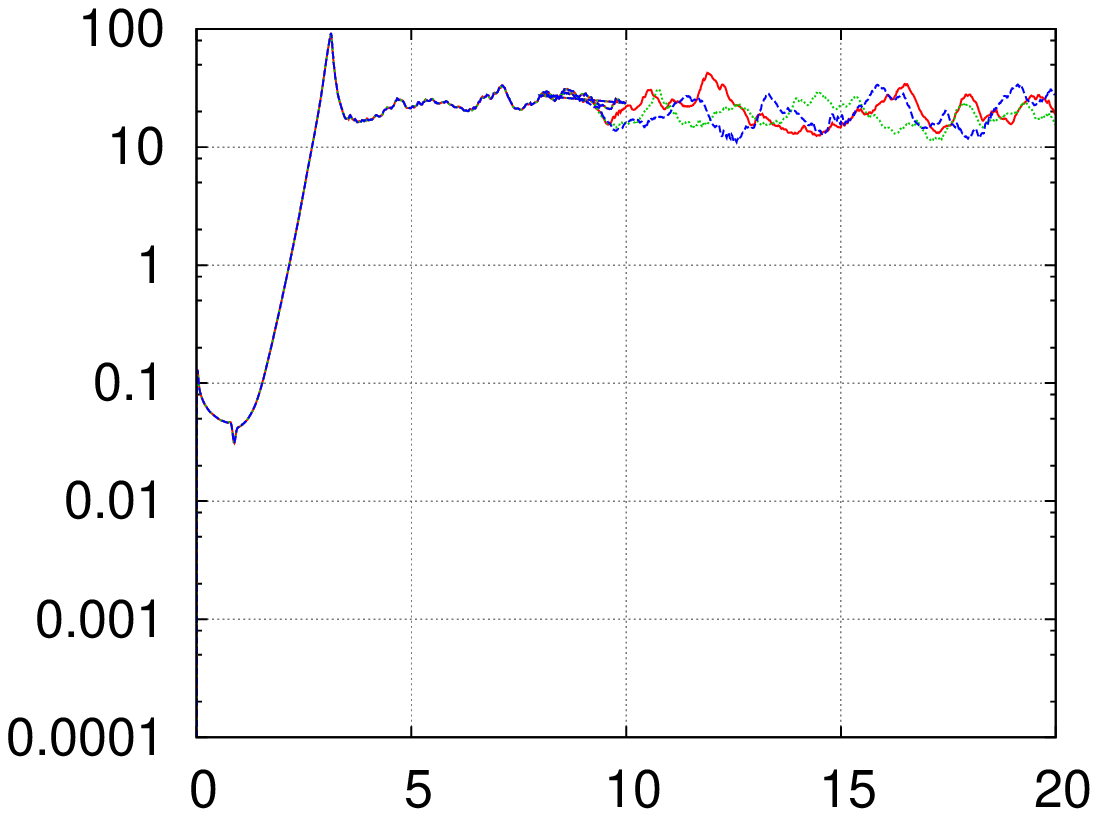}} \\
      \parbox{0.4cm}{(b) \\ $\bigtriangleup$} &
      \raisebox{-4.02cm}{\includegraphics[width=7cm]{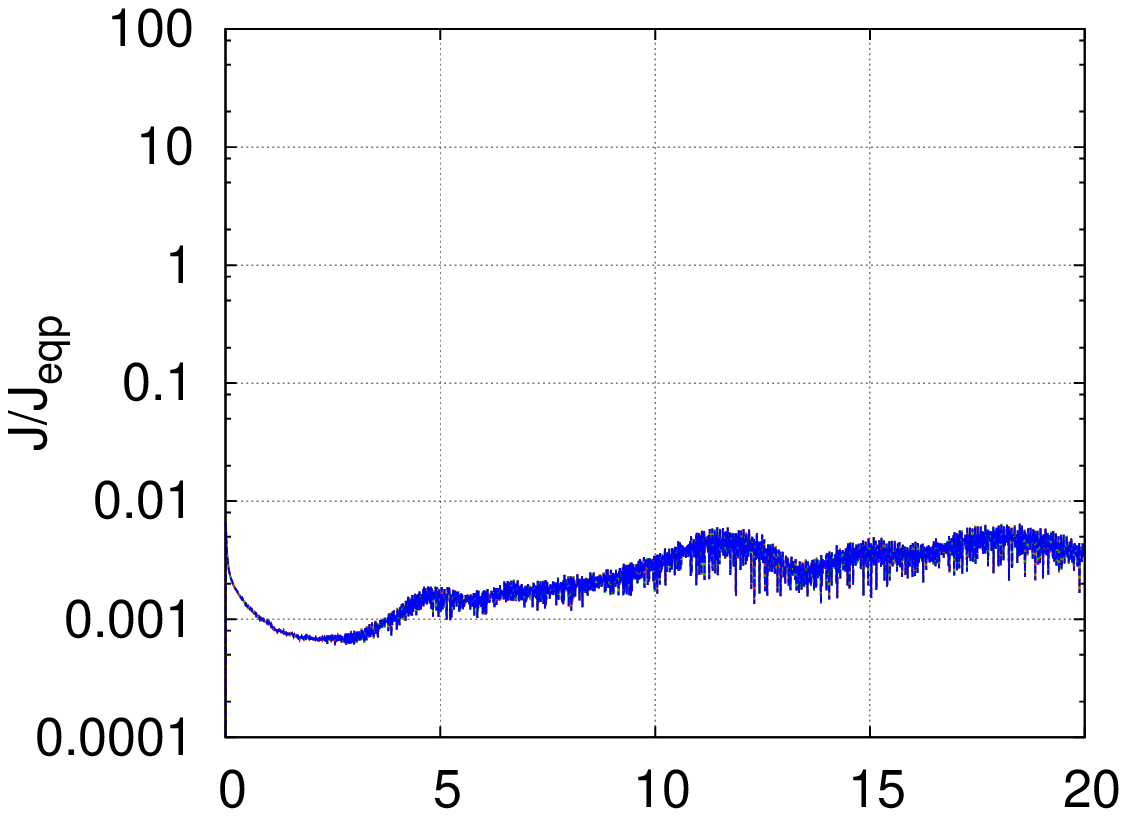}} & 
      \raisebox{-4.02cm}{\includegraphics[width=7cm]{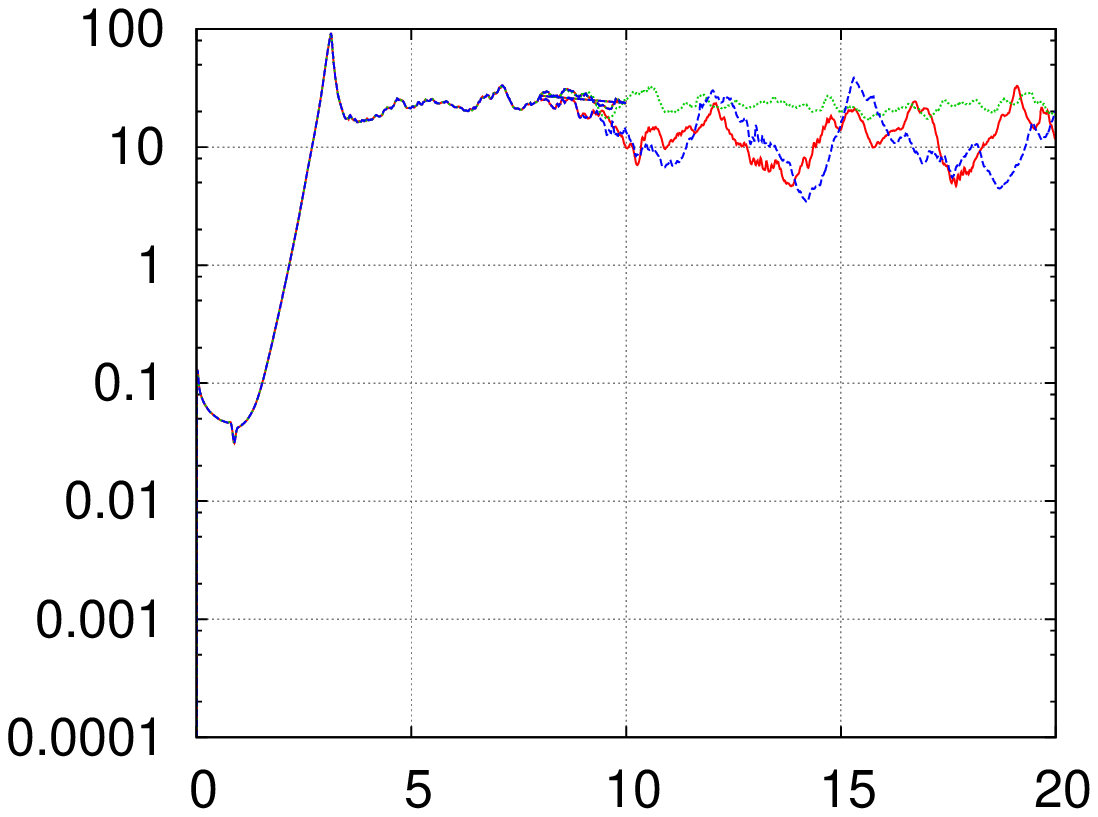}} \\
      \parbox{0.4cm}{(c) \\ $\times$} &
      \raisebox{-4.02cm}{\includegraphics[width=7cm]{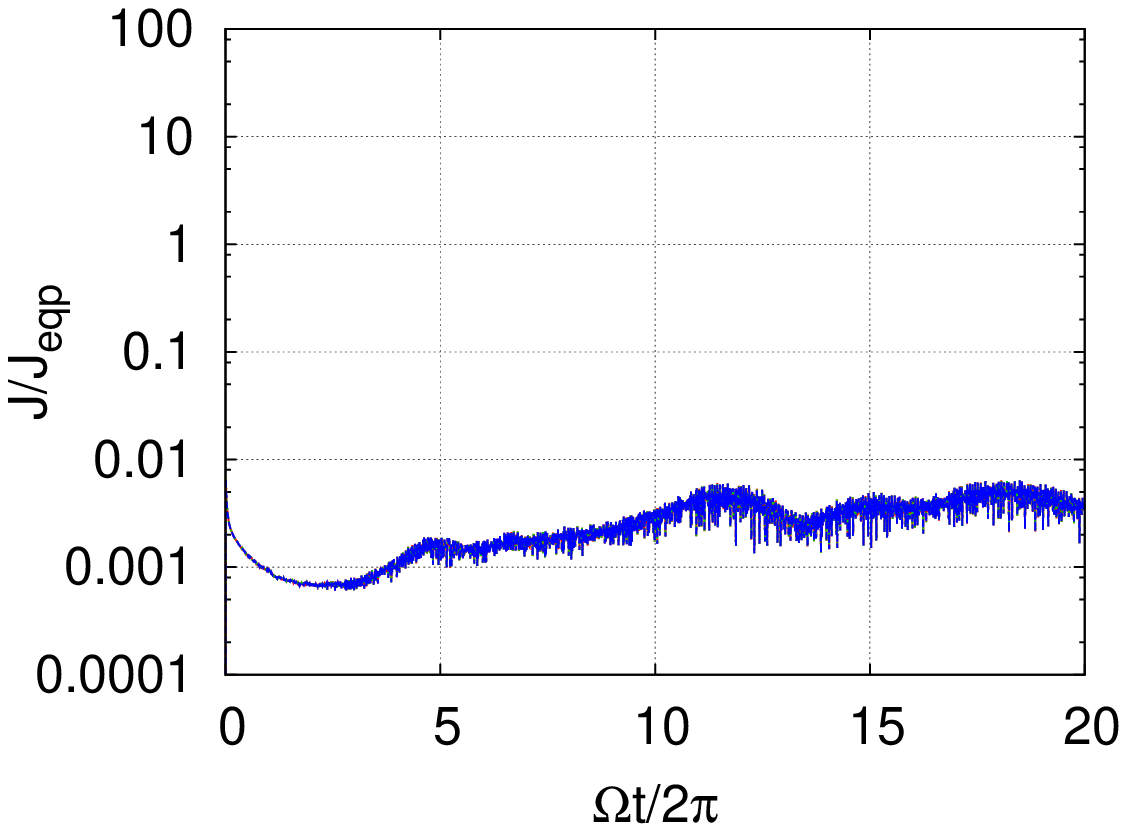}} & 
      \raisebox{-4.02cm}{\includegraphics[width=7cm]{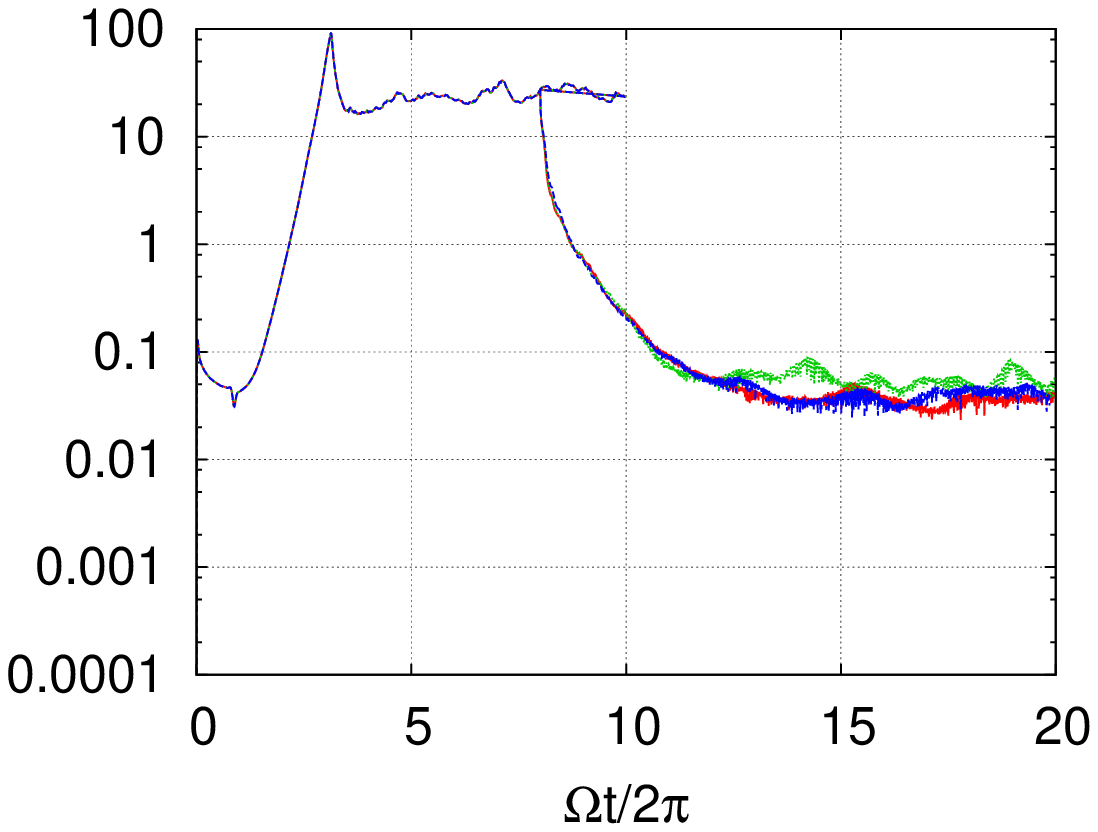}} \\
    \end{tabular}
  \end{center}
  \caption {
    \addspan {
      Dependence of the MRI behavior on resistivity models. 
      The development of electric current over time in the 
      simulations using different resistivity models 
      {\tt fid}(  Eqs. (\ref{eqNonlinearEfield}) (\ref{eqNonlinearMagDiff})),
      {\tt p2} (Eq. \ref{eqNonlinearPTwo})
      and {\tt p4} (Eq. \ref{eqNonlinearPFour}),
      for three sets of parameters
      (a) $\beta=400$, $R_M=0.6$, $\criticalCurrent/\equipartitionCurrent = 1$ ,
      (b) $\beta=400$, $R_M=0.2$, $\criticalCurrent/\equipartitionCurrent = 1$ and 
      (c) $\beta=400$, $R_M=0.2$, $\criticalCurrent/\equipartitionCurrent = 10$ ;
      which are typical parameter sets for 
      (a) $\bigcirc$       active zone,
      (b) $\bigtriangleup$ sustained zone, and
      (c) $\times$         dead zone, respectively.
    }
  }\label{figChangeModel}
\end{figure*}

\addspan{ To study possible influences of the numerical resolutions,
  we have performed 
  simulations using the different numerical resolution
    $N_x = 32,48,64$(fiducial) and $80$ while keeping the aspect ratio $N_x : N_y : N_z = 1:2:1$, 
    for three different set of parameters
    $(\beta, R_M,
  \criticalCurrent/\equipartitionCurrent)$ = (400,0.6,1), (400,0.2,1)
  and (400,0.2,1).
  Figure \ref{figConvergence} summarizes the convergence tests.
  Within these sets of parameters, we observe that our classification does not depend
  on the numerical resolution.
}

\addspan{ 
  We have also studied if our result depends on the model of
  nonlinear Ohm's law. In addition to our fiducial({\tt fid}) model,
  Eq. (\ref{eqNonlinearMagDiff}), we
  have studied two smoothly transitting models, Eqs. 
(\ref{eqNonlinearPTwo}) and
(\ref{eqNonlinearPFour}). To distinguish the three models see Figure \ref{figNonlinearModels}.
}

\addspan{
  Figure \ref{figChangeModel} summarizes the time evolution of the current density
  in the simulations with the three typical sets of parameters
  $(\beta, R_M, \criticalCurrent/\equipartitionCurrent)$ =
  (400,0.6,1), (400,0.2,1) and (400,0.2,1). 
  Within the regime we have tested, the hysteresis behavior does not depend on the detail
  of the non-linear resistivity models.
}

From Figures \ref{figSMRIsurvey} and   \ref{figSMRIsurvey2} we can see the following condition for the $\bigtriangleup$sustained zone:
\begin{eqnarray}
  \frac {\criticalCurrent}{\equipartitionCurrent} \frac{1}{\magneticReynolds} &\leq& \sustainFactor \label{eqWHBalanceModel} ,
\end{eqnarray}
where $\sustainFactor$ is a proportionality constant that satisfies
$\sustainFactor \simeq 5-15$ for $400 \leq \beta \leq 1600$, and
$\sustainFactor \simeq 15-50$ for $\beta \geq 3200$.
Hereafter, we interpret this in terms of the {\it work-heat balance} 
in the resistive MRI.

%

{
We also remark that within parameter regions that is in active zone, $\magneticReynolds < 1$, and
with larger $\criticalCurrent$,
we observed the large-amplitude time-variability of physical quantities such as magnetic fields due to 
repeated growth and reconnection of the channel solutions. This phenomenon is reported by
\citet{fleming_effect_2000}.
}

\subsection{Interpretation of The Simulation Results}\label{sectionInterpretation}
In this section, we show that Equation (\ref{eqWHBalanceModel}) can be
understood as a condition of the balance between the magnetic energy
dissipated by Joule heating per unit volume ($\jouleHeating$) and the
work done by shearing motion per unit volume ($\shearWork$).

Let us define $\lhs$ as the left hand side of Equation (\ref{eqWHBalanceModel}):
\begin{eqnarray}
  \lhs \equiv  \frac {\criticalCurrent}{\equipartitionCurrent} \frac{1}{\magneticReynolds}.
\end{eqnarray}
The condition for self-sustained MRI is $\lhs < \sustainFactor$, which is to be explained in this section.

First, substitute $\magneticReynolds \equiv {\alfvenVelocityZ}^2 / \magDiff_0 \Omega $;
\begin{eqnarray}
  \lhs = \frac { \criticalCurrent}{\equipartitionCurrent} \frac{\magDiff_0 \Omega}{ {\alfvenVelocityZ}^2} 
  \label{eqInterpret01}
\end{eqnarray}

Next, when the MRI is active, 
$ \langle \current^2
\rangle^{1/2}  \simeq \saturationJF  \equipartitionCurrent$
where $\saturationJF$ is of the order of $\saturationJFValue$.
Since this average current strength lies in super-critical regime of Ohm's law
($\current > \criticalCurrent$),
the electric field is $ \criticalEfield =
4 \pi c^{-2}\linearMagDiff \criticalCurrent $ as modeled in
Equation (\ref{eqNonlinearEfield}) and (\ref{eqNonlinearMagDiff}).
Therefore, Joule heating per unit volume is estimated as
\begin{eqnarray}
  \jouleHeating 
  &=&  \criticalEfield \cdot \langle \current^2 \rangle^{1/2} 
  \nonumber \\
  &=&  \criticalEfield \cdot \saturationJF  \equipartitionCurrent \nonumber \\
  & = & 4 \pi   \saturationJF  c^{-2}\linearMagDiff \criticalCurrent \equipartitionCurrent 
  \label{eqInterpret02-proto} .
\end{eqnarray}

{ To explain the work-heat balance qualitatively, this
  estimate needs correction due to the high space variability of
  current field under the discharge conditions. We introduce
  $\fillingFactor$, the filling factor, the ratio of the volume that
  contributes to the Joule heating to the total volume. Formally,
  $\fillingFactor$ is defined as the ratio between volume averages of
  the actual Joule heat generated and the Joule heat estimated by this
  method:
\begin{eqnarray}
  \fillingFactor &\equiv& 
  \frac{\langle \Ecom J  \rangle}
       {  \criticalEfield \cdot \langle \current^2 \rangle^{1/2} }
  \nonumber \\
  &=&\frac
  {\langle f\left(J\right) \current \rangle} 
  {\langle \current^2 \rangle^{1/2}} ,
\end{eqnarray}
where
\begin{eqnarray}
  f\left(\current\right) = \left\{
  \begin{array}{ccl}
    1                                 & \mathrm{if} & \current > \criticalCurrent \\
    \frac{\current}{\criticalCurrent} & \mathrm{if} & \current < \criticalCurrent 
  \end{array}
  \right. .
\end{eqnarray}
}

Using this $\fillingFactor$, Equation (\ref{eqInterpret02-proto}) is rewritten as:
\begin{eqnarray}
  \jouleHeating 
  & \equiv & 4 \pi  \fillingFactor \saturationJF  c^{-2}\linearMagDiff \criticalCurrent \equipartitionCurrent 
  \label{eqInterpret02} .
\end{eqnarray}

Substituting $\magDiff_0$ in (\ref{eqInterpret01}) with (\ref{eqInterpret02}) gives
\begin{eqnarray}
  \lhs &=&\frac {c^2 \jouleHeating \Omega}{ 4 \pi  \fillingFactor \saturationJF \equipartitionCurrent^2  {\alfvenVelocityZ}^2}  \nonumber , \\
  \jouleHeating &=&
  \frac { 4 \pi  \fillingFactor \saturationJF  \equipartitionCurrent^2  {\alfvenVelocityZ}^2 \lhs}  {c^2  \Omega}.
  \label{eqInterpret03}  
\end{eqnarray}

Substituting $\alfvenVelocityZ$ with Equation (\ref{eqPlasmaBeta2}) and then
$\soundSpeed$ with (\ref{eqScaleHeight}),
$\equipartitionCurrent$ with Equation (\ref{eqUnitCurrent}) and then
$\equipartitionMagnet$ with Equation (\ref{eqUnitMagnet}) gives
\begin{eqnarray}
  \jouleHeating &=&
  \frac {4 \fillingFactor \saturationJF \pressure_0  \Omega \lhs}{ \beta} \label{eqInterpret04}.
\end{eqnarray}

On the other hand,
\begin{eqnarray}
  \shearWork &\equiv& \frac{3}{2} \Omega \stressTensor \nonumber \\
  &=& \frac{3}{2} \alpha \Omega \pressure \label{eqShearWork} ,
\end{eqnarray}
where $\alpha$ is \citet{shakura_black_1973} 's  $\alpha$ parameter.
Substituting Equation (\ref{eqShearWork}) into Equation (\ref{eqInterpret04}), one obtains
\begin{eqnarray}
  \jouleHeating &=& 
  \frac {8  \fillingFactor \saturationJF \lhs}{3  \alpha  \beta} \shearWork \label{eqInterpret}.
\end{eqnarray}
For the MRI to sustain itself by the discharge process,
the Joule heating
$\jouleHeating$ needs to be equal or smaller than
$\shearWork$:
\begin{eqnarray}
  \jouleHeating &\simleq& \shearWork.
\end{eqnarray}
Therefore, we have the following constraint on the left hand side of Equation (\ref{eqInterpret01}):
\begin{eqnarray}
  \lhs &=&
  \frac{3  \alpha  \beta}{8  \fillingFactor \saturationJF}
  \frac{\jouleHeating}{\shearWork} \nonumber,   \\
  \frac {\criticalCurrent}{\equipartitionCurrent} \frac{1}{\magneticReynolds} 
  &\simleq&
  \frac{3  \alpha  \beta}{8  \fillingFactor \saturationJF} \equiv \sustainFactor\left(\beta\right). \label{eqLHSExplain}
\end{eqnarray}
Thus, the work-heat balance poses an upper limit on the product of
${\criticalCurrent}/{\equipartitionCurrent}$ and
${1}/{\magneticReynolds}$, provided that  $\fillingFactor$,
$\saturationJF$ and $\alpha$ are constants that do not depend on
${\criticalCurrent}$ and ${\magneticReynolds}$, but only on $\beta$.
This explains the inverse-proportional relations observed in
Figures \ref{figSMRIsurvey} and   \ref{figSMRIsurvey2}. The $\sustainFactor(\beta)$ calculated with
this interpretation using the experimental data are in Table
\ref{tableSustainFactor}.

\begin{table*}
  \begin{center}
    \begin{tabular}{c|cccc|c}
      $\beta$ &
      $\overline{\alpha}$ &
      $\displaystyle \overline{\left\langle \frac{ B^2}{8 \pi P_0}  \right\rangle}$    &         
      $\overline{\saturationJF}$ &
      $\overline{\fillingFactor}$  & 
      $\sustainFactor(\beta) $ \\
      \hline
      400     & $0.176 \pm 0.036$  & $0.222 \pm 0.040$ & $14.5 \pm 1.7$   & $0.252 \pm 0.001$ & $7.19 \pm 0.81$\\
      800     & $0.143 \pm 0.035$  & $0.206 \pm 0.057$ & $17.4 \pm 2.7$   & $0.259 \pm 0.001$ & $9.33 \pm 1.03$\\
      1600    & $0.104 \pm 0.027$  & $0.166 \pm 0.047$ & $17.4 \pm 1.8$   & $0.265 \pm 0.002$ & $13.3 \pm 2.3$\\
      3200    & $0.0523 \pm 0.0206$& $0.0916 \pm 0.0411$ & $12.2 \pm 3.9$ & $0.262 \pm 0.008$ & $19.0 \pm 3.9$\\
      6400    & $0.0236 \pm 0.0041$& $0.0386 \pm 0.0072$ & $9.58 \pm 1.04$& $0.263 \pm 0.003$ & $22.2 \pm 1.6$\\
      12800   & $0.0182 \pm 0.0032$& $0.0300 \pm 0.0060$ & $9.84 \pm 0.88$& $0.270 \pm 0.000$ & $32.6 \pm 2.8$\\
      25600   & $0.0185 \pm 0.0063$& $0.0315 \pm 0.0128$ & $10.7 \pm 1.5$ & $0.275 \pm 0.001$ & $58.5 \pm 12.4$
    \end{tabular}
  \end{center}
  \caption{
    The $\sustainFactor(\beta)$ calculated from the experimental data. We first calculated the
    time and space averaged quantities 
    $\overline{\alpha}$ and
    $\overline{\saturationJF}$ 
    for each runs. Then for the ensemble of runs,
    we calculated the means and the standard deviations of the quantities.
    The ensemble constitutes of runs
    that (1) belong to $\bigtriangleup$sustained zone,
    (2) are restarted runs ($t = 16\pi/\Omega, 18\pi/\Omega, 20\pi/\Omega$)
    so that they are magnetorotationally unstable, and
    (3) have the largest product
    ${\criticalCurrent}/{\equipartitionCurrent}
    \cdot {1}/{\magneticReynolds}$ so that they face the 
    $\bigtriangleup$sustained zone - 
    $\times$dead zone boundary.
  }\label{tableSustainFactor}
\end{table*}

{
We can further simplify Equation (\ref{eqLHSExplain}) 
by using the saturation predictor proposed by \HGB. The proposed predictors read
\begin{eqnarray}
  \alpha P &=& 0.61 \pm 0.06 \overline{\left\langle \frac{B^2}{8 \pi} \right\rangle} ,\\
  \overline{\left\langle \frac{B^2}{8 \pi} \right\rangle}
  &=& \left(1.21 \pm 0.29\right) \cdot 2 \pi \sqrt{\frac{16}{15} \cdot\frac{2}{\beta}} P_0 .\\  
\end{eqnarray}
Using this, Equation (\ref{eqLHSExplain}) is rewritten as follows:
\begin{eqnarray}
  \frac {\criticalCurrent}{\equipartitionCurrent} \frac{1}{\magneticReynolds} 
  &\simleq&  \sustainFactor\left(\beta\right) \nonumber \\
  &\simeq&\left(2.54 \pm 0.66\right)
  \frac{\beta^{1/2}}{\fillingFactor \saturationJF} .
\end{eqnarray}
By ignoring the dependence of $\fillingFactor$ and $\saturationJF$ on $\beta$,
 we assume 
$\overline{\fillingFactor}= 0.264 \pm 0.007$ and
$\overline{\saturationJF} = 13.1 \pm 3.1$.
This further simplifies the 
Equation (\ref{eqLHSExplain}) as:
\begin{eqnarray}
  \sustainFactor\left(\beta\right)
  & \simeq&\left(0.74 \pm 0.26\right)
                {\beta^{1/2}}.
\end{eqnarray}
This is in agreement with our experimental data (Table \ref{tableSustainFactor}) within factor of 2.

\section{Distribution of The Three MRI Zones within The Protoplanetary Disks}

\subsection{Protoplanetary Disk Model}
{
In the previous section, we performed the shearing-box simulations of MRI with nonlinear Ohm's
law, and found the three classes of MRI behavior; we named them active, dead and sustained zones.
We also found the condition for the MRI to be self-sustained. In this section, we apply the findings
to the global model of the protoplanetary disks and analyze how they are divided into the three zones.
}

We use Minimum-Mass Solar Nebula (MMSN) model introduced by
\citet{hayashi_formation_1985} as the fiducial disk model;
\begin{eqnarray}
  \surfaceDensity = \surfaceDensityFactor \fiducialSurfaceDensity \rOverAU ^ {-\fiducialSurfaceDensityPower} 
  \label{eqSurfaceDensityModel} , \\
  \temperature = \fiducialTemperature \rOverAU ^ {-\fiducialTemperaturePower} .
  \label{eqTemperatureModel} 
\end{eqnarray}
Here, $\fiducialSurfaceDensity = 1.7 \times 10^3 \unit{g\ cm^{-3}}$
and $\fiducialTemperature = 2.8 \times 10^2 \unit{K}$ are the surface
density and the temperature at 1AU, respectively.
$\surfaceDensityFactor$ is the nondimensional surface density
parameter.  Fiducial value for the surface density power index is
$\fiducialSurfaceDensityPower = 3/2$.  Since we assume the isotropic
equation of state (EOS), the ratio of specific heats $\gamma = 1$ in
our model, and the thermal velocities for gas molecules and electrons
are
\begin{eqnarray}
  \soundSpeed\left(r\right) &=& \sqrt{\frac{  \BoltzmannConstant  \temperature }{\fiducialMolecularWeight  \MassOfHydrogen}} , \\
  \electronThermalVelocity\left(r\right) &=& \sqrt{\frac{  \BoltzmannConstant  \temperature }{\MassOfElectron}} ,
\end{eqnarray}
respectively, where $\fiducialMolecularWeight$ is the mean molecular weight of the gas.

Since we neglect the self gravity of the disk, the disk is in Kepler rotation and its orbital angular velocity is
\begin{eqnarray}
  \orbitalOmega\left(r\right) &=& \sqrt{\frac{\GravitationalConstant \starMass }{ r^3 }} .
\end{eqnarray}

From the equilibrium between vertical pressure gradient and vertical component of the stellar gravity, the disk density
and pressure distributions are:
\begin{eqnarray}
  \density\left(r,z\right) &=& \frac{\surfaceDensity}{ \sqrt{2 \pi} \scaleHeight}
  \exp\left( {\frac{- z^2 } {2 \scaleHeight^2}}\right) , \\
  \pressure\left(r,z\right) &=& \frac{\density  \BoltzmannConstant  \temperature }{\fiducialMolecularWeight  \MassOfHydrogen},
\end{eqnarray}
where $\scaleHeight$ is the definition of the disk scale-height in this paper c.f. Equation (\ref{eqScaleHeight}).

For simplicity, we assume that the every dust particle to be solid sphere of
the equal radius $\dustRadius$ and density
$\dustMaterialDensity$. The mass $\dustMass$ and geometrical cross
section $\dustCrossSection$ of the dust particle are
\begin{eqnarray}
  \dustMass &=& \frac {4 \pi}{3} \dustMaterialDensity \dustRadius^3
  \label{eqDustMass}, \\
  \dustCrossSection &=& \pi \dustRadius^2,
\end{eqnarray}
respectively. The fiducial values are
$\dustRadius = 0.1 \unit{\mu m}$ and
$\dustMaterialDensity = 3 \unit {g \ cm^{-3}}$.

Using this $\dustMass$ and dust-to-gas density ratio $\dustGasRatio = 0.01$, 
the number densities of dust and gas component are:
\begin{eqnarray}
  \gasNumberDensity\left(r,z\right) &=& \frac{\density}{\fiducialMolecularWeight \MassOfHydrogen}, \\
  \dustNumberDensity\left(r,z\right) &=& \frac{\dustGasRatio \hspace{1mm} \density}{\dustMass} 
  \label{eqDustNumberDensity} . \\
\end{eqnarray}

\subsection{Ionization Processes}
Methods for calculating the charge equilibrium of the dust-plasma in the protoplanetary disks has been studied
\citep{umebayashi_recombination_1980,umebayashi_effects_2009,fujii_fast_2011}.
Among those we use \ZUMIZUMI's method because of its numerical efficiency and generality.
First, the gas column density above ($\cdAbove$) and below ($\cdBelow$) the certain coordinate $(r,z)$ in the disk are
\begin{eqnarray}
  \cdAbove\left(r,z\right) &=& \int_z^{\infty}  \density \mathit {dz} \nonumber \\
  &=& \frac {\surfaceDensity} {2} \left[1 - \erf\left(\frac {z}{\sqrt 2 \scaleHeight}\right) \right], \\
  \cdBelow\left(r,z\right) &=& \int_{-\infty}^z \density \mathit {dz} \nonumber \\
  &=& \frac {\surfaceDensity} {2} \left[1 + \erf\left(\frac {z}{\sqrt 2 \scaleHeight}\right) \right],
\end{eqnarray}
respectively, where
\begin{eqnarray}
  \erf\left(x\right) \equiv \frac{2}{\sqrt \pi} \int_x^{\infty} e^{-t^2} \mathit {dt}
\end{eqnarray}
is the error function.

According to \SMUN, the effective ionization rate in the disk is
\begin{eqnarray}
  \ionizationRate\left(r,z\right) &\approx& \frac {\ionizationRateCR} {2}
  \left\{ \exp\left(-\frac{\cdAbove }{ \attenuationLengthCR}\right) \right. \nonumber \\
  & + & \left. \exp\left(-\frac{\cdBelow }{\attenuationLengthCR}\right) \right\}  
            + \ionizationRateRA,
\end{eqnarray}
where $ \attenuationLengthCR = 96 \unit{g\ cm^{-2}}$,
$\ionizationRateCR = 1.0\times 10^{-17} \unit{s^{-1}}$, and
$\ionizationRateRA = 6.9\times 10^{-23} \unit{s^{-1}}$.

Using this, \ZUMIZUMI's nondimensional parameter $\Theta$ is calculated as
\begin{eqnarray}
  \Theta = \frac{\ionizationRate \gasNumberDensity \ElementaryCharge^2}
        {\stickingI \soundSpeed  \dustCrossSection \dustRadius \dustNumberDensity^2 
          \BoltzmannConstant \temperature},\label{eqBigTheta}
\end{eqnarray}

and $\Gamma$ is defined as the solution of the equation
\begin{eqnarray}
  \frac{1}{1+\Gamma} - \left(\frac{\stickingI}{\stickingE}\sqrt{\frac{\MassOfElectron}{\fiducialMolecularWeight \MassOfHydrogen}} 
  \exp \Gamma + \frac {\Gamma} {\Theta}\right) = 0. \label{eqGamma}
\end{eqnarray}
Once Equation (\ref{eqGamma}) is numerically solved for $\Gamma$, we can calculate
the number density of ions and electrons, $\ionNumberDensity, \electronNumberDensity$, 
as well as the root mean square of the charge per dust particle, $\sqrt{\dustChargeZSq} e$, as:
\begin{eqnarray}
  \ionNumberDensity\left(r,z\right) &=& \frac {\ionizationRate \gasNumberDensity}  {
    \stickingI  \soundSpeed \dustCrossSection \dustNumberDensity \left(1+\Gamma\right)}, \\
  \electronNumberDensity\left(r,z\right) &=& \frac{\ionizationRate \gasNumberDensity \exp \Gamma } {
    \stickingE \electronThermalVelocity \dustCrossSection \dustNumberDensity }, \\
  \dustChargeZSq &=& \left(\frac{\Gamma \dustRadius }{\lambda}\right)^2 + 
  \frac{1+\Gamma}{2+\Gamma}\frac{\dustRadius}{\lambda}\\
  \mathrm{where}~\lambda &=&\frac{\ElementaryCharge^2}{\BoltzmannConstant \temperature},\nonumber
\end{eqnarray}
We assume sticking probabilities $s_i =1 $ and $ s_e = 0.3$
as assumed by \ZUMIZUMI. 

Next, we estimate the plasma conductivity using the method of \SMUN.
The rate coefficient for the collision between the neutrals and the ions is
\begin{eqnarray}
  \sigmaVi = 2.41 \pi \left(\frac{\alpha \ElementaryCharge^2 }{\gasMass}\right)^{\frac 1 2}.
\end{eqnarray}
We use $\alpha = 7.66 \times 10^{-25} \unit{cm^{3}}$ as an averaged polarizability.
$\sigmaVe$ is the rate coefficient for collision between neutrals and
electron, whose form is given in \SMUN. The rate coefficient for dust particles
is
\begin{eqnarray}
  \sigmaVd = \frac{4\pi}{3}\dustRadius^2 \soundSpeed \label{eqDustSigmaV}.
\end{eqnarray}
This expression is valid as long as
$\dustRadius$ is much smaller than the mean free path of the gas molecules.

{
With these, the magnetic diffusivity is calculated component-wise:
\begin{eqnarray}
  \magDiffE&=& \frac{c^2 \MassOfElectron \gasNumberDensity \sigmaVe}
           {4 \pi \ElementaryCharge^2  \electronNumberDensity} 
           ,  \\
  \magDiffI&=& \frac{c^2 \gasMass \gasNumberDensity \sigmaVi}
           {4 \pi \ElementaryCharge^2  \ionNumberDensity} 
           , \\
  \magDiffD&=& \frac{c^2 \gasMass \gasNumberDensity \sigmaVd}
           {4 \pi \dustChargeZSq \ElementaryCharge^2  \dustNumberDensity} 
           , \\
  \linearMagDiff &=&  \left({\magDiffE}^{-1} +   {\magDiffI}^{-1} +  {\magDiffD}^{-1}\right)^{-1} 
\end{eqnarray}
}
The value of $\criticalEfield$ is set by the condition that the
kinetic energy of the electrons accelerated by the electric field is
large enough to initiate the electron avalanche.
\begin{eqnarray}
  \ionizationEnergy & = &\dpCoef \hspace{1mm} \ElementaryCharge \hspace{1mm} \criticalEfield  \hspace{1mm} l_{\mathrm {mfp}} ; \nonumber\\
  \criticalEfield &=& \frac{\ionizationEnergy}{\dpCoef \hspace{1mm} \ElementaryCharge \hspace{1mm} l_{\mathrm {mfp}}} \nonumber \\
  &=& \frac{\ionizationEnergy \gasNumberDensity \sigmaVe}{\dpCoef e \electronThermalVelocity}.
  \label{eqEcritDP}
\end{eqnarray}
Here, $\dpCoef = 0.43 \sqrt{\gasMass / \MassOfElectron}$ 
is the coefficient for average energy of electrons in weakly ionized plasma
(\ISSS),
and $l_{\mathrm {mfp}} = \electronThermalVelocity/( \gasNumberDensity \sigmaVe)$ is the mean
free path of electrons.
For ionization energy we use the value for a hydrogen molecule
$\ionizationEnergy = 15.4\unit{eV}$.

With this, the critical current is
\begin{eqnarray}
  \criticalCurrent = \frac{\SpeedOfLight^2}{4 \pi \linearMagDiff} \criticalEfield .
\end{eqnarray}

{
Note that the discharge electric field, Equation (\ref{eqEcritDP}) is calculated
using the strong electric field limit of the electron distribution function,
 while we used the weak field limit formulae for charge distributions,
Equations (\ref{eqBigTheta})-(\ref{eqDustSigmaV}). In this paper we adopt this
treatment for simplicity. A more consistent treatment
will be the topic of another paper in preparation.
}

\begin{figure*}
  \begin{center}
    \hspace{-1cm}
    \includegraphics[angle=270,width=9cm]{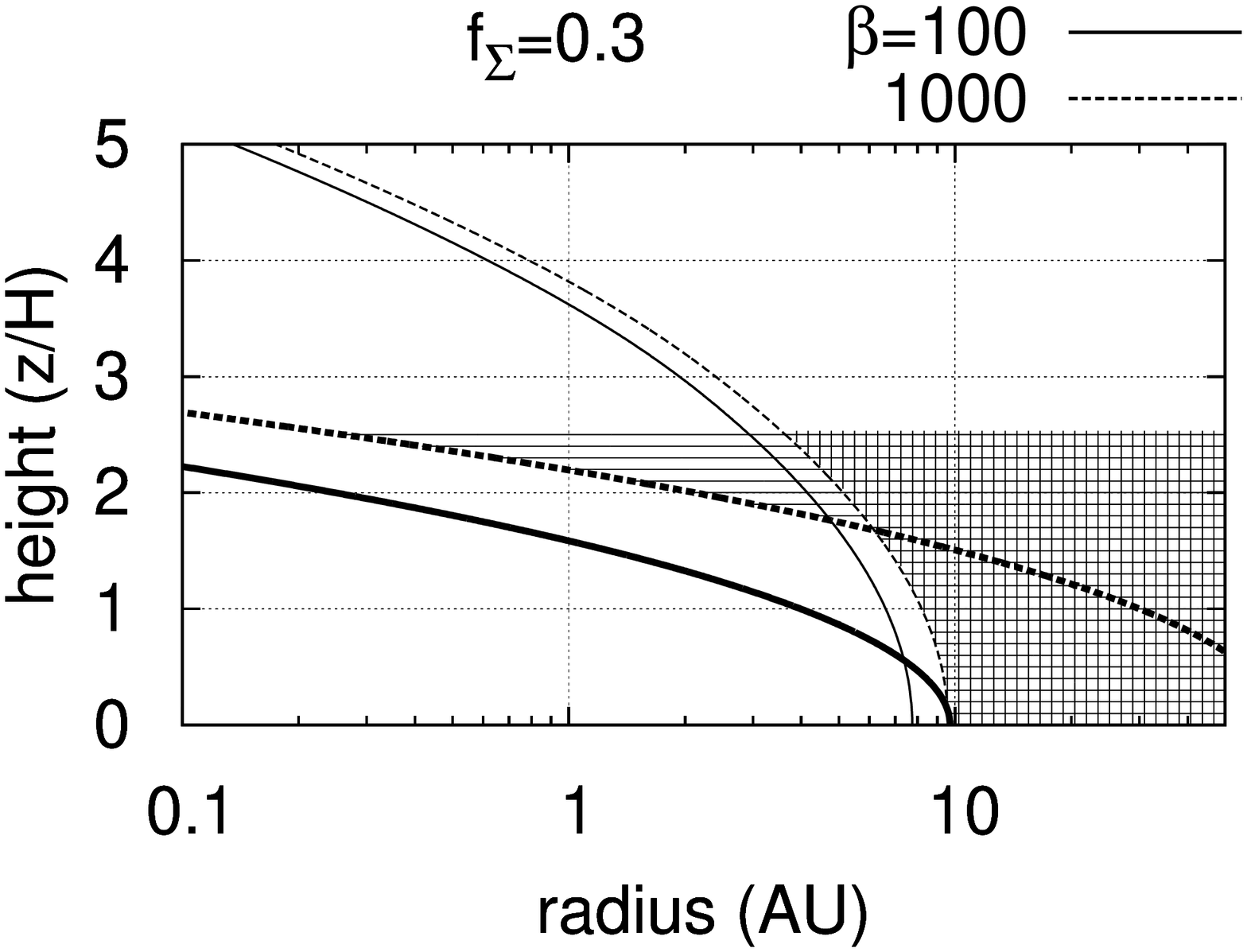}\hspace{-1cm}
    \includegraphics[angle=270,width=9cm]{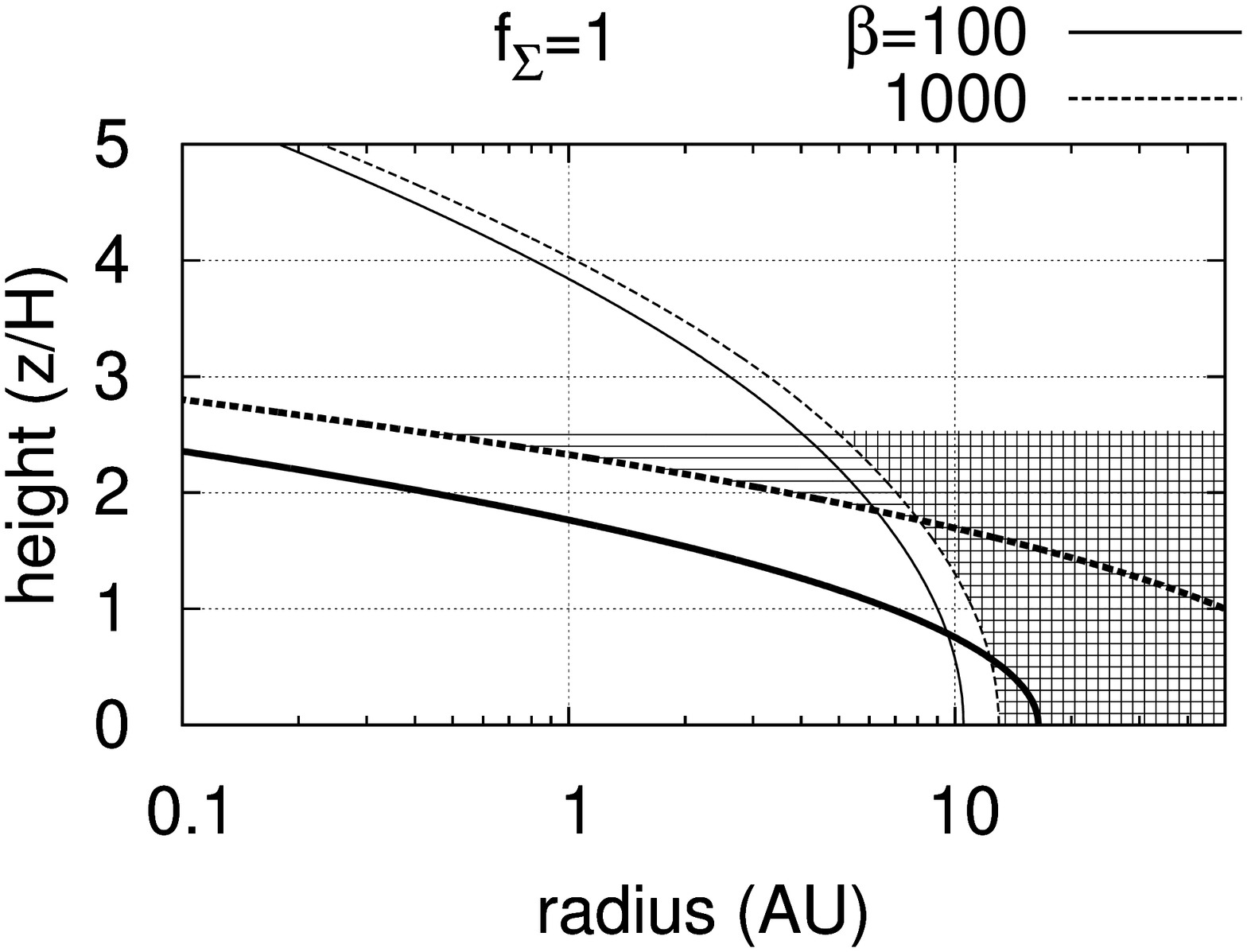}\hspace{-1cm}
    
    \hspace{-1cm}
    \includegraphics[angle=270,width=9cm]{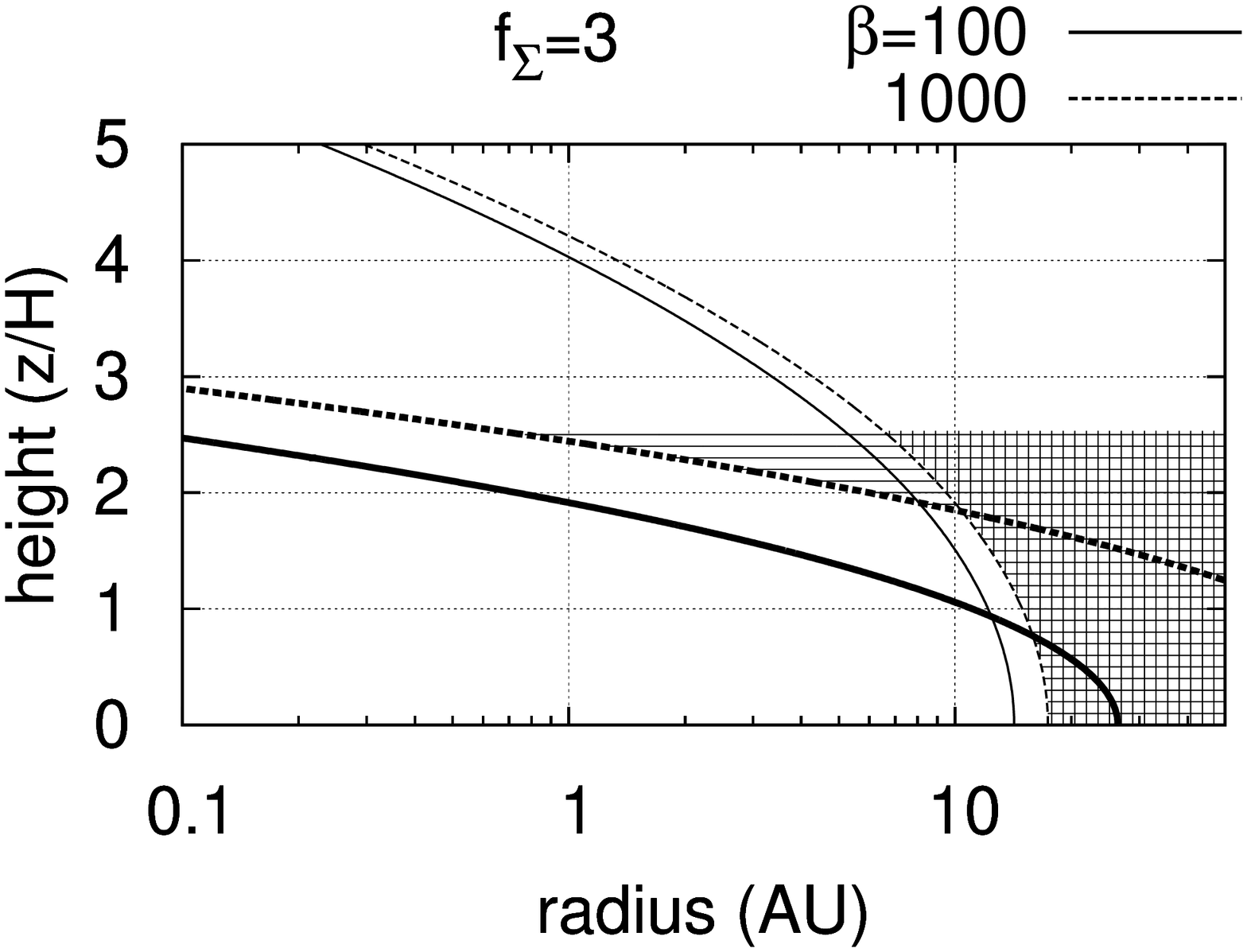}\hspace{-1cm}
    \includegraphics[angle=270,width=9cm]{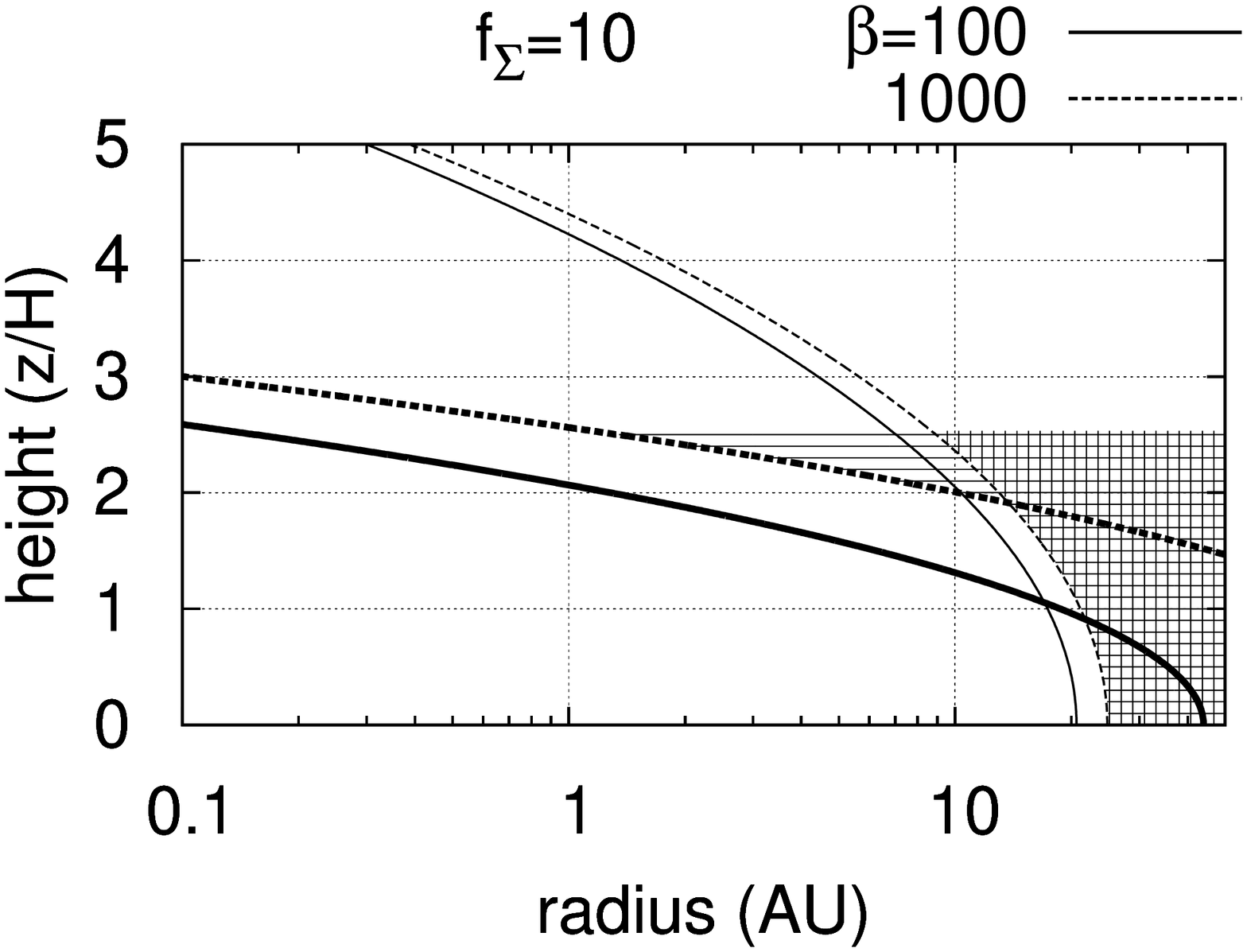}\hspace{-1cm}

  \end{center}
  \caption { Unstable regions in the protoplanetary disks. The
    thin solid and thin dashed curves represent $\lambdaRes/{\sqrt 2}H
    = 1$ for the cases of the magnetic field strength $\beta=100$ and
    $1000$, respectively, inside of which is dead zone if the MRI
    self-sustainment is not taken into account (\SMUN).  The regions
    above the thick solid and thick dashed curves satisfies
    Equation (\ref{eqWHBalanceModel}) for $\beta=100$ and $1000$,
    respectively, and are sustained zones according to the work-heat
    balance model.  We compare the unstable region predicted by
    \SMUN\ to that predicted by our model for $\beta = 1000$.  The
    unstable regions according to \SMUN's and our model are marked by
    vertical and horizontal stripes, respectively.}
  \label{figureRegionSD}
\end{figure*}

\begin{figure*}
  \begin{center}
    \hspace{-1cm}
    \includegraphics[angle=270,width=9cm]{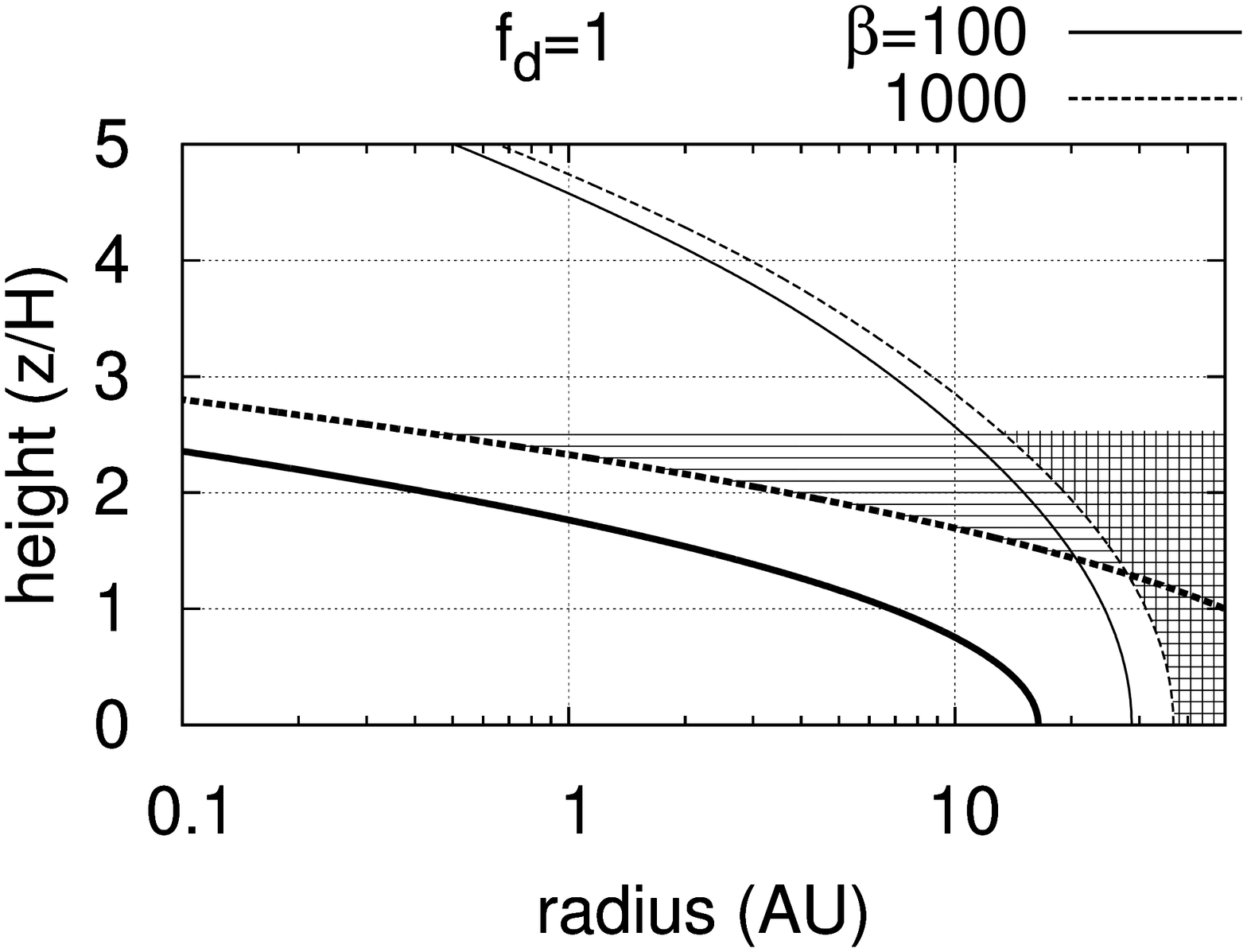}\hspace{-1cm}
    \includegraphics[angle=270,width=9cm]{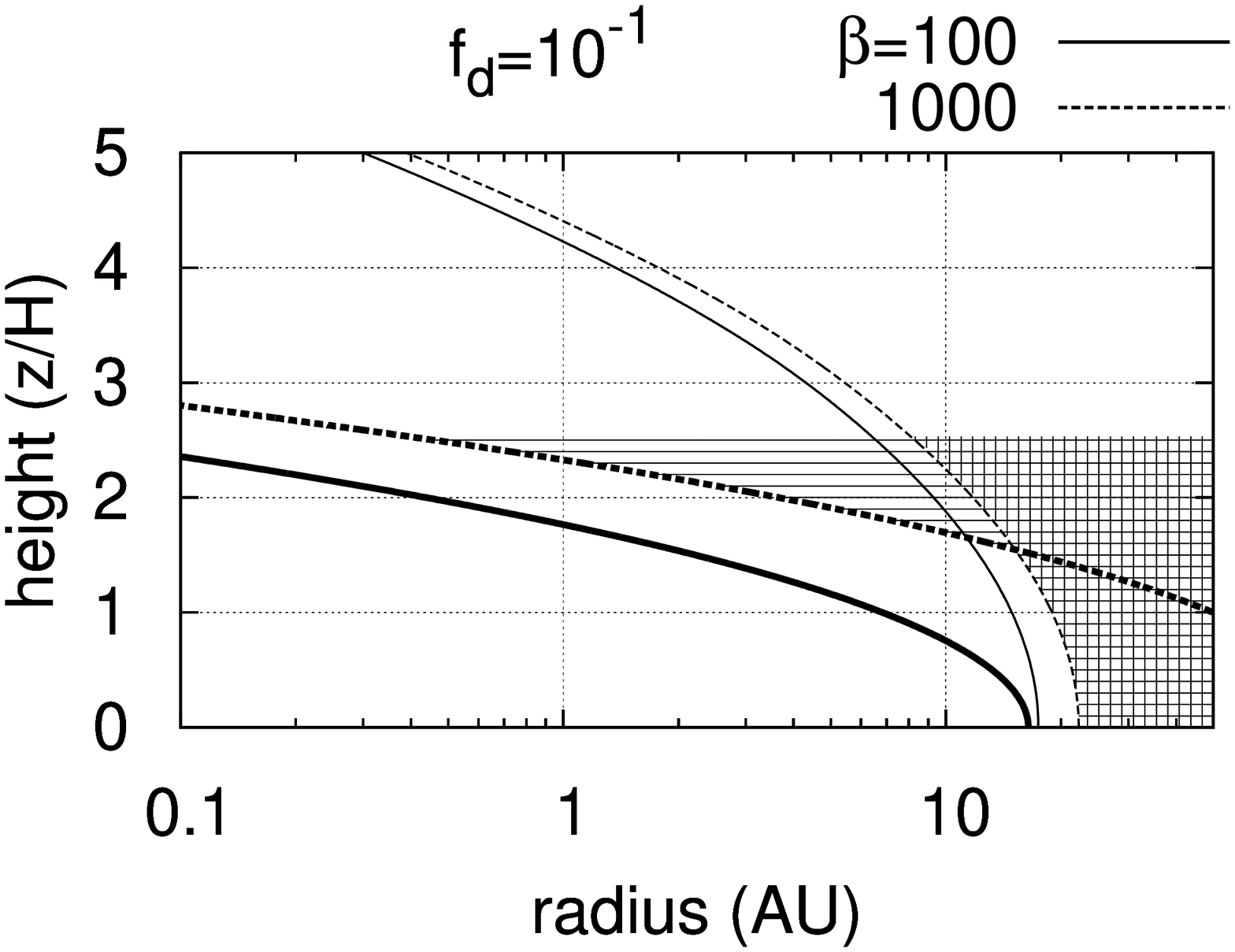}\hspace{-1cm}

    \hspace{-1cm}
    \includegraphics[angle=270,width=9cm]{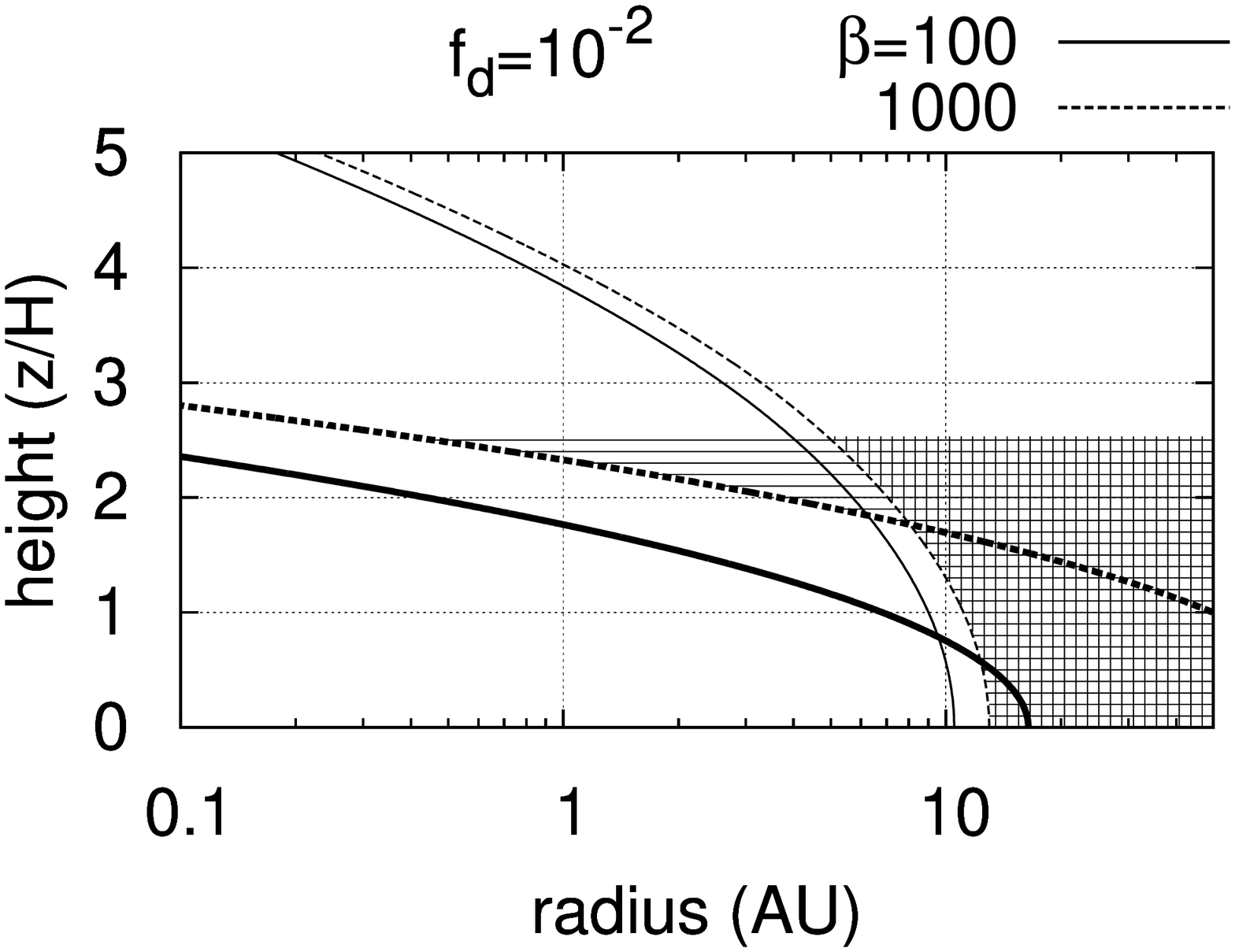}\hspace{-1cm}
    \includegraphics[angle=270,width=9cm]{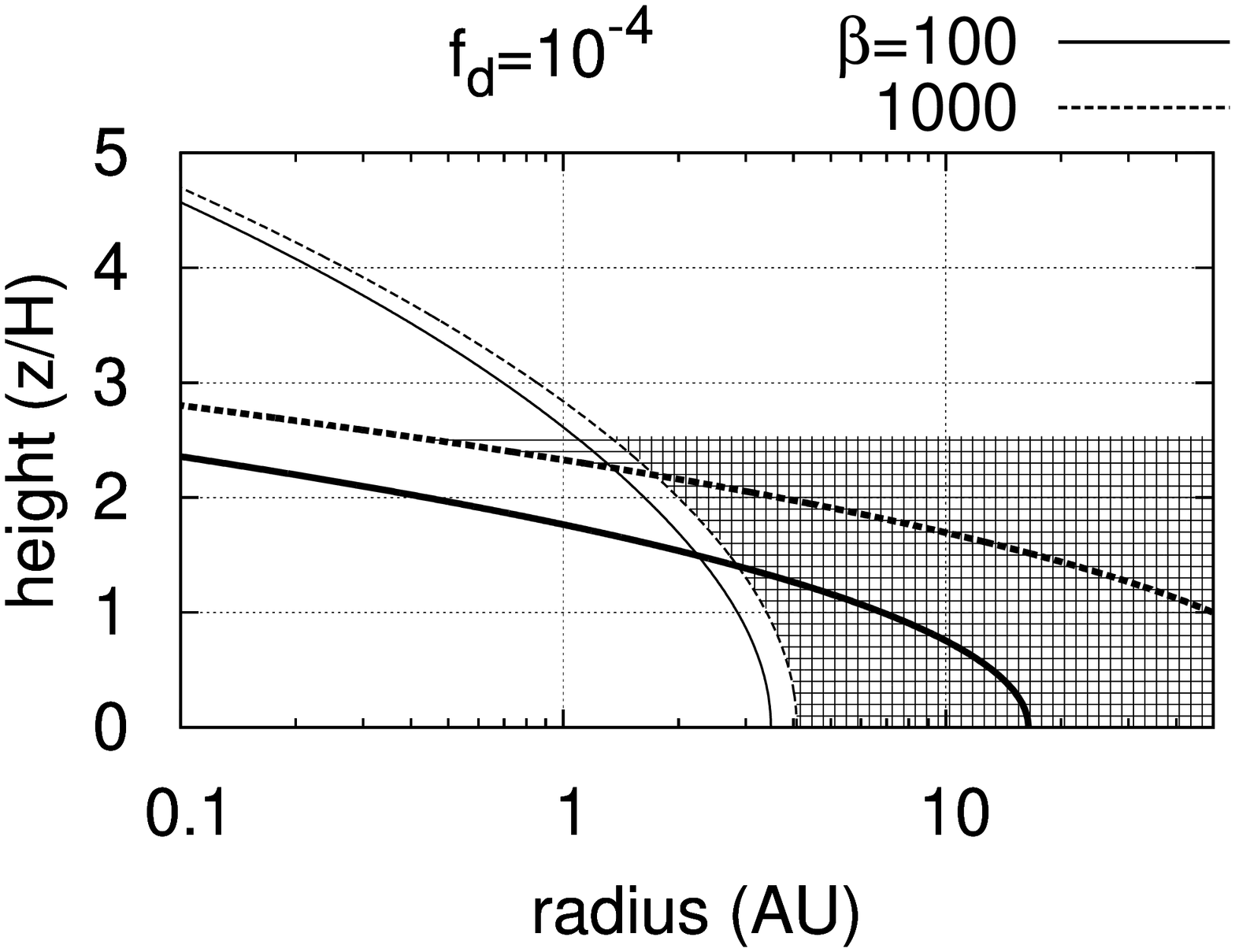}\hspace{-1cm}
  \end{center}
  \caption { Unstable regions for different dust-to-gas ratio.
    The MRI-unstable region according to (\SMUN) and our model are marked by
    vertical and horizontal stripes, respectively, for $\beta = 1000$.
  }      \label{figureRegionDEP}
\end{figure*}

\begin{figure*}
  \begin{center}
    \hspace{-1cm}
    \includegraphics[angle=270,width=9cm]{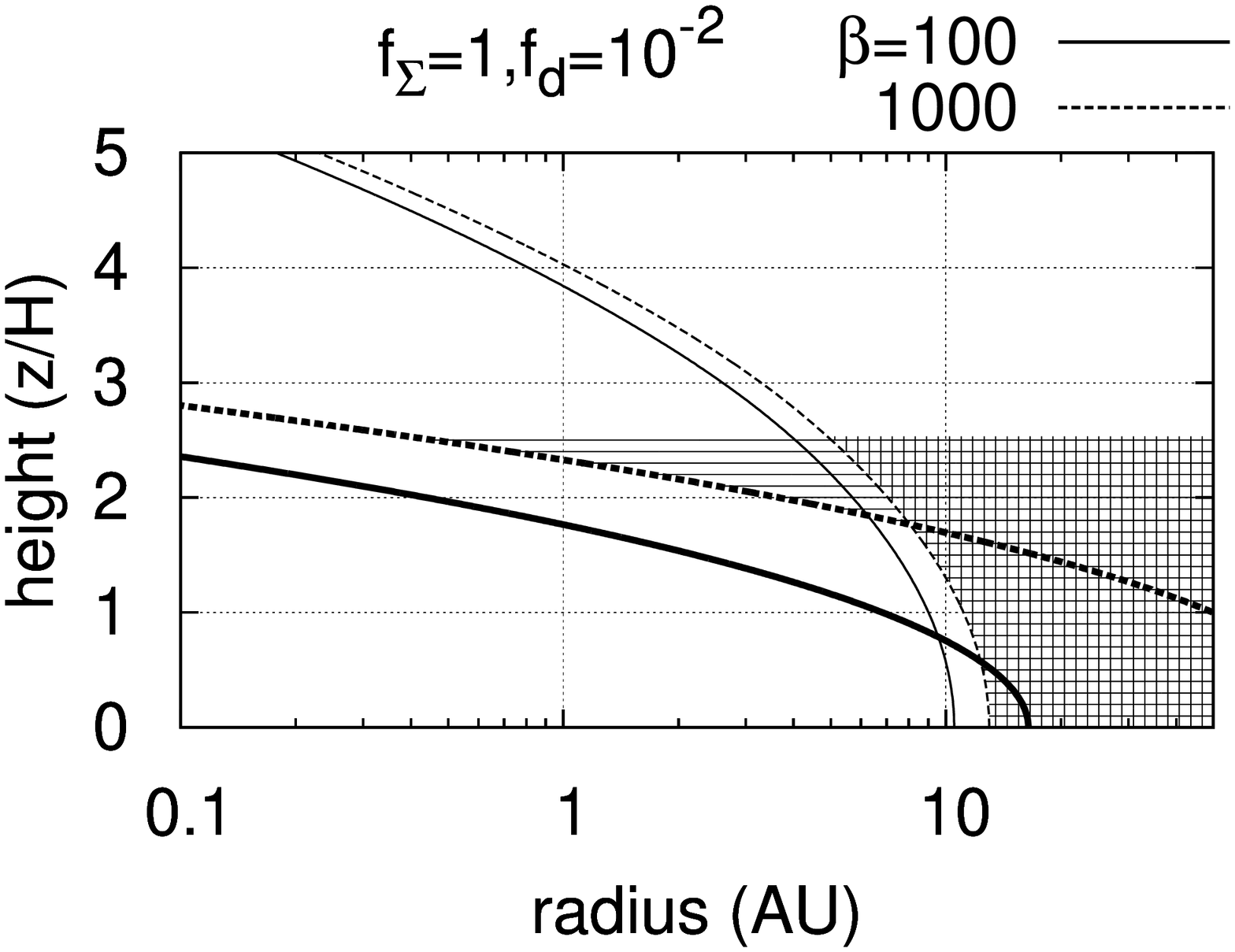}\hspace{-1cm}
    \includegraphics[angle=270,width=9cm]{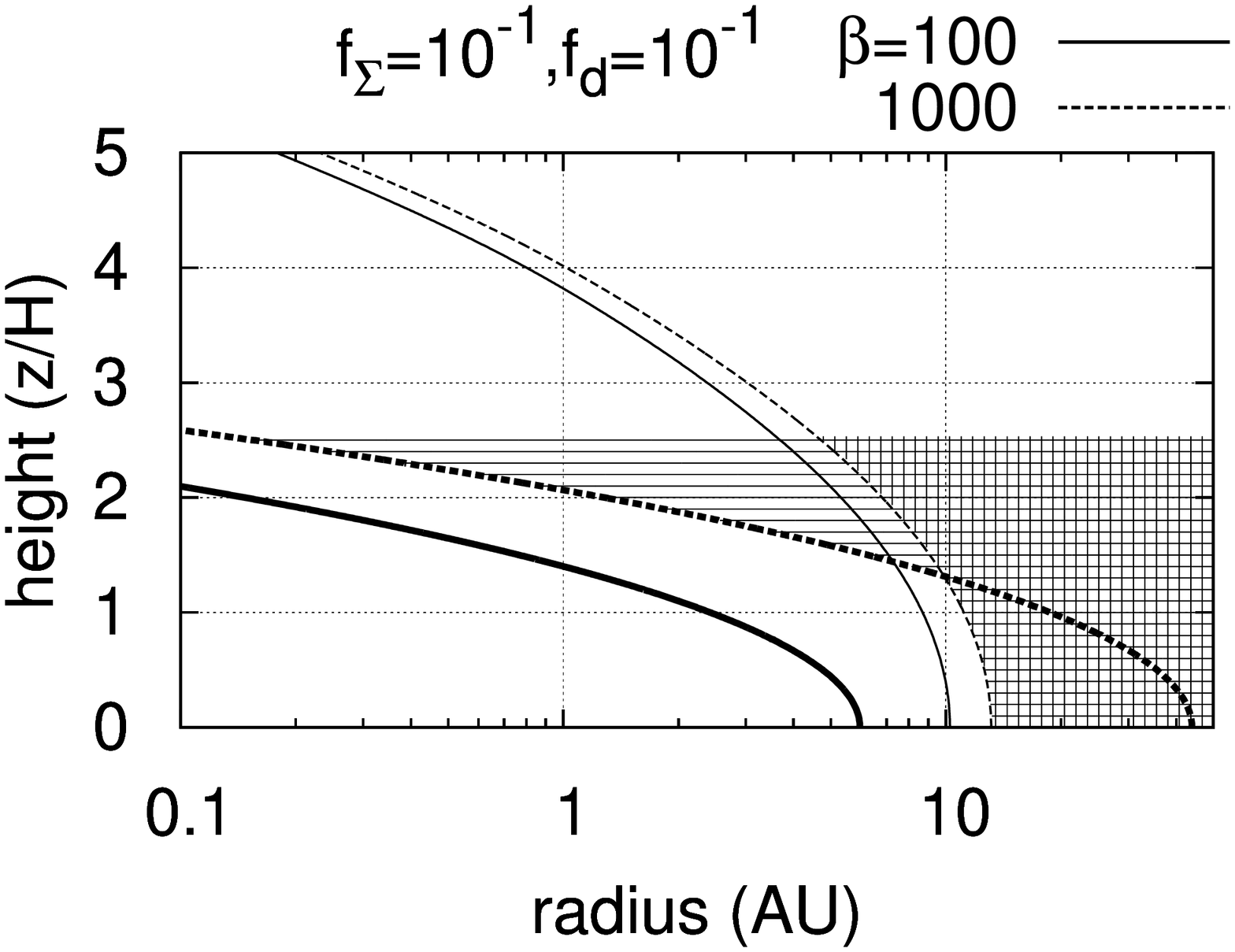}\hspace{-1cm}

    \hspace{-1cm}
    \includegraphics[angle=270,width=9cm]{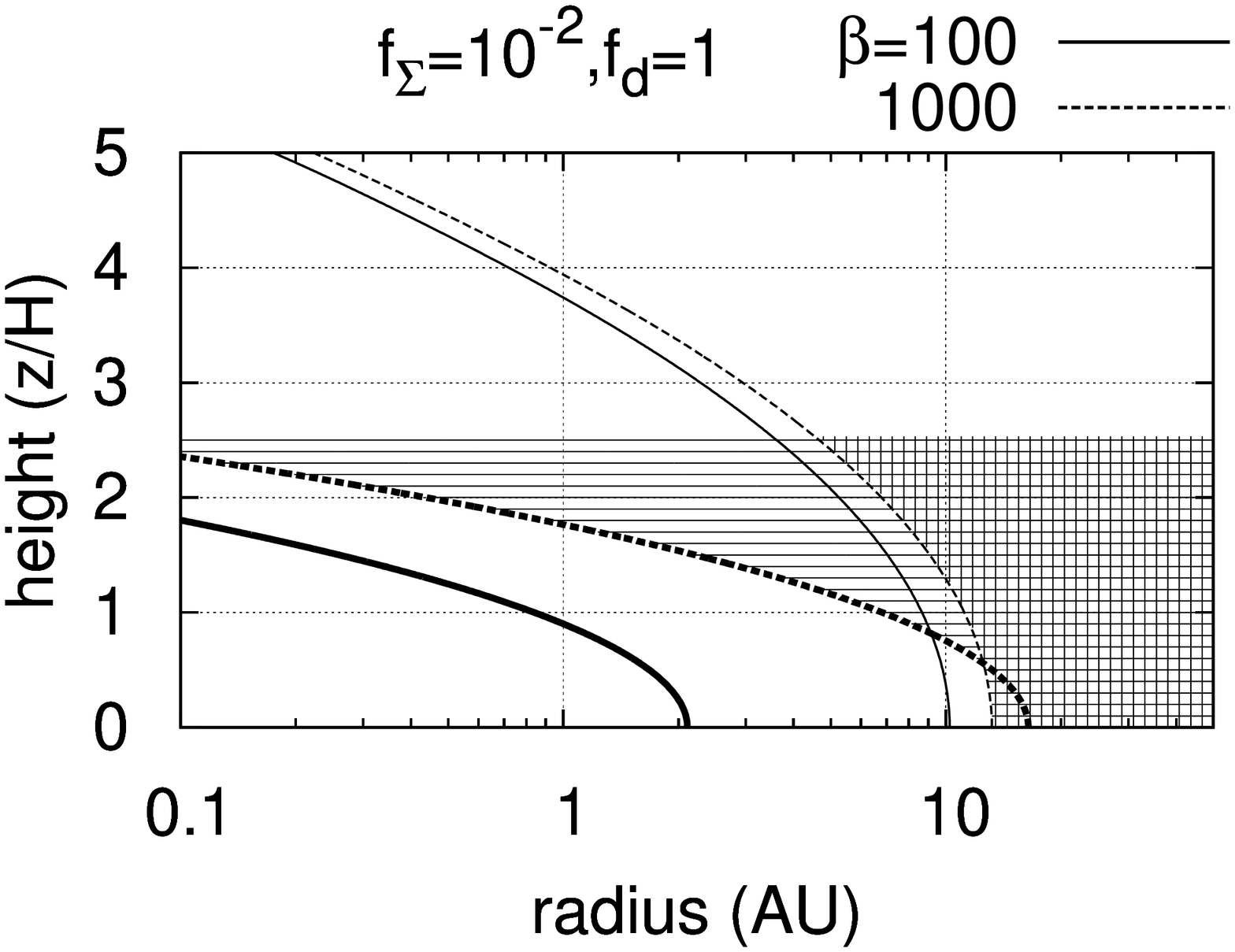}\hspace{-1cm}
    \includegraphics[angle=270,width=9cm]{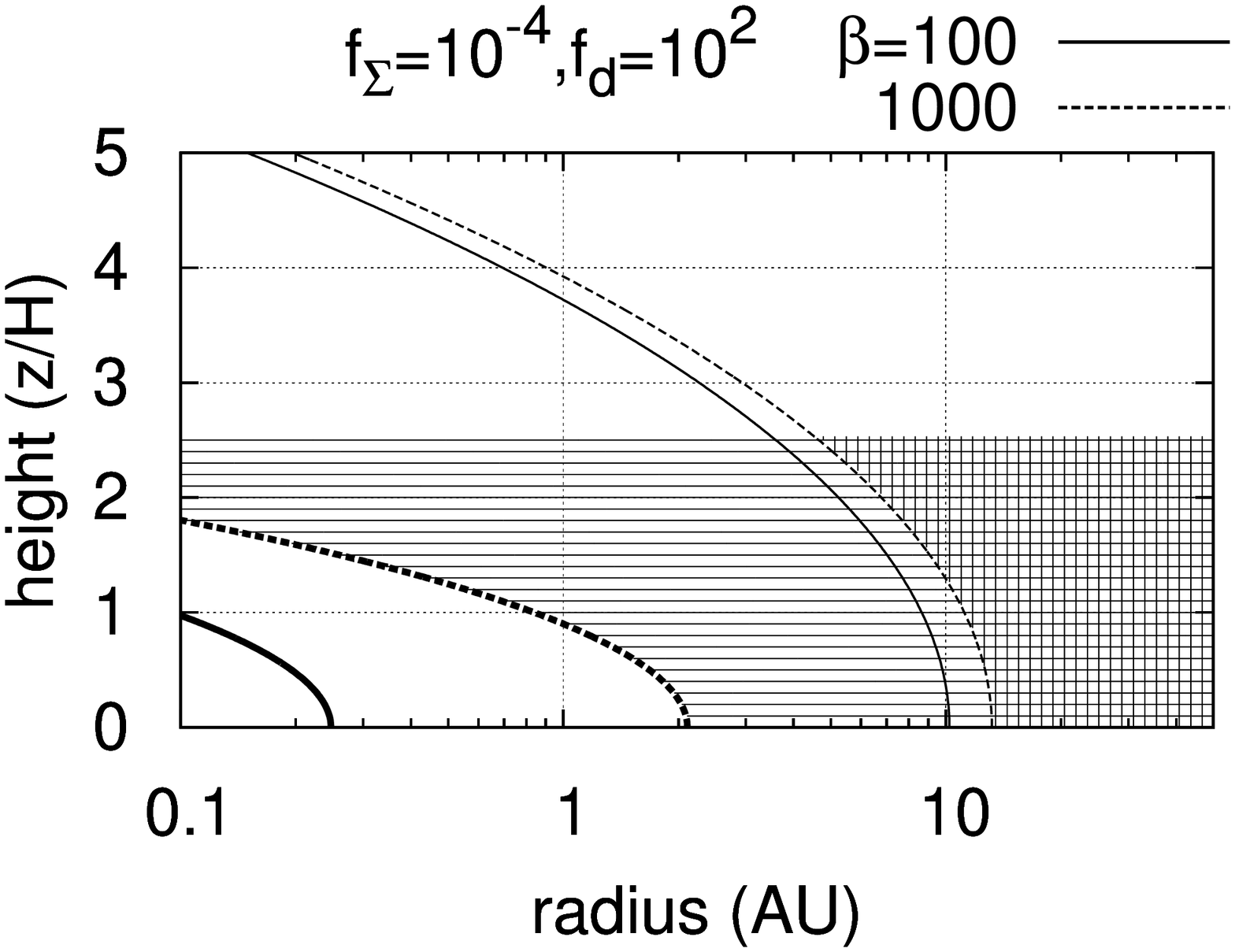}\hspace{-1cm}
  \end{center}
  \caption { Change of unstable regions as the gas density of the disk decreases, 
    while the dust density is kept constant. 
    The MRI-unstable region according to (\SMUN) and our model are marked by
    vertical and horizontal stripes, respectively, for $\beta = 1000$.
  }      \label{figureRegionTAKERU}
\end{figure*}

\subsection{The Self-Sustained MRI in Global Disk Models}

Now we study the distribution of active, sustained and dead zones in the protoplanetary disk models.
\SMUN ~ gives the condition for 
 MRI unstable region as follows:
\begin{eqnarray}
  \frac{2\pi\alfvenVelocityZ}{\Omega} \leq \sqrt{2} H\bigwedge\frac{2\pi\magDiff}{\alfvenVelocityZ} \leq \sqrt{2} H 
  \label{eqSanoMiyama} .
\end{eqnarray}

Combination of  this with work-heat balance model Equation (\ref{eqWHBalanceModel}) gives the following conditions
for active, sustained and dead zones, respectively:
\begin{eqnarray}
  & &\hspace{-0.7cm}\frac{2\pi\alfvenVelocityZ}{\Omega}  \leq \sqrt{2} H
  \bigwedge  \frac{2\pi\magDiff}{\alfvenVelocityZ} \leq \sqrt{2} H \label{eqGlobalActiveZone} \\
  & &\hspace{-0.7cm} \frac{2\pi\alfvenVelocityZ}{\Omega}  \leq \sqrt{2} H
  \bigwedge  \frac{2\pi\magDiff}{\alfvenVelocityZ} > \sqrt{2} H 
  \bigwedge  \frac {\criticalCurrent}{\equipartitionCurrent} \frac{1}{\magneticReynolds} \leq \sustainFactor
  \nonumber \\  \label{eqGlobalSustainedZone} \\
  & &\hspace{-0.7cm} \frac{2\pi\alfvenVelocityZ}{\Omega}  > \sqrt{2} H
  \bigvee \left( \frac{2\pi\magDiff}{\alfvenVelocityZ} > \sqrt{2} H 
  \bigwedge  \frac {\criticalCurrent}{\equipartitionCurrent} \frac{1}{\magneticReynolds} > \sustainFactor  \right)
  \nonumber  \label{eqGlobalDeadZone} \\ 
\end{eqnarray}

Using these Equations (\ref{eqGlobalActiveZone})-(\ref{eqGlobalDeadZone}), we plot
the active, sustained and dead zones for various global disk model
Equations (\ref{eqSurfaceDensityModel})-(\ref{eqTemperatureModel}).

First, Figure \ref{figureRegionSD} shows the unstable zones for varying disk surface density,
$\surfaceDensityFactor = 0.3$, 
$\surfaceDensityFactor = 1$ (the fiducial model), 
$\surfaceDensityFactor = 3$, and
$\surfaceDensityFactor = 10$.
In this figure and following figures, the thick curves are the boundary of the  
 work-heat balance model Equation (\ref{eqWHBalanceModel}), while the thin curves are 
 the boundary of the instability condition in the resistive limit, i.e. the second condition in  Equation  
(\ref{eqSanoMiyama}). The solid and dashed curves correspond to the plasma beta at the mid-plane
 $\beta = 100$ and  $\beta = 1000$, respectively.
The active zones are marked by meshes, and the sustained zones are
marked by horizontal stripes.



Figure \ref{figureRegionDEP} shows the active and sustained zones for 
dust-to-gas ratio
$\dustGasRatio = 1$,
$\dustGasRatio = 0.1$,
$\dustGasRatio = 0.01$ (the fiducial model), and
$\dustGasRatio = 10^{-4}$.
The work-heat balance condition 
 Equation (\ref{eqWHBalanceModel}) is not affected 
by changing the dust properties such as
dust-to-gas ratio $\dustGasRatio$ or dust size $\dustRadius$.
 One can understand this by rewriting  the condition
Equation (\ref{eqInterpret01}) in the following form:
\begin{eqnarray}
  \frac{ \criticalEfield  \Omega}{4 \pi c^{-2} \equipartitionCurrent {\alfvenVelocityZ}^2} \leq \sustainFactor.
\end{eqnarray}
This form does not include a term affected by the dust properties, such as the magnetic diffusivity. On the other hand,
$\criticalEfield$ is inversely proportional to the electron mean free path $l_{\mathrm {mfp}}$, and is proportional to the gas number density.

In Figure \ref{figureRegionTAKERU}
we study  the evolution of the active and sustained zones 
as the gas density becomes lower while the dust density is kept constant. 
We change the set
$(\surfaceDensityFactor, \dustGasRatio)$ from
$(1, 0.01)$ (the fiducial model) to
$(0.1, 0.1)$,
$(0.01, 1)$, and
$(10^{-4}, 100)$. In
Figure \ref{figureRegionTAKERU}
zones are marked for $\beta = 1000$.
The midplane of the disk between the radii $2\unit{AU} - 20\unit{AU}$ becomes the
sustained zone  as the gas density becomes $10^{-4}$ times the fiducial model.

\section{Conclusions and Discussions}

{
By performing numerical simulations of MHD with nonlinear Ohm's law 
of the three-dimensional local disks,
we
found hysteresis behavior for certain diffusivity model: If we start
from the laminar-flow initial conditions with small seed fluctuations,
the MRI does not activate because of the diffusivity and the flow remains
laminar; on the other hand, if we take MRI-turbulent state from an
ideal-MHD simulation as initial conditions, MRI remains active under
the same diffusivity model.
We have surveyed in three-parameter space
($\beta$, $\criticalCurrent$, $\magneticReynolds$)
in search for the regions the self-sustained MRI in the context of \ISSS\
takes place.
We found the condition, the work-heat balance model, for this
hysteresis behavior to take place. The model is $\jouleHeating \simleq \shearWork$,
where $\jouleHeating$ is the magnetic energy dissipated by Joule
heating per unit volume and $\shearWork$ is the work done by background shearing
motion per unit volume. This leads to the proportionality relation
$\frac {\criticalCurrent}{\equipartitionCurrent} \frac{1}{\magneticReynolds} \leq \sustainFactor$.
}

{
\ISSS\ concluded that the the energy supply from the \addspan{shearing motion}
should be $\sim 30$ times greater than the energy needed to supplying the enough ionization for the MRI,
and predicted the entire disk to be active. However,
applying the work-heat balance model to various protoplanetary disk models,
we have found that in most of the models, the sustained zone is above $z/H>2 - 3$.
}

{
We conclude that in the fiducial protoplanetary disks environment the
Joule heating (which has been neglected in \ISSS) becomes the dominant
energy dissipation channel and constrains the self-sustainment of MRI,
and the midplane of the disk remain dead.  However, the gas of the
disk dissipates 
(\citet{alexander_photoevaporation_2006, alexander_photoevaporation_2006-1,suzuki_protoplanetary_2010})
with observed timescale of $10^6-10^7$ years
(\citet{cieza_spitzer_2007,hernandez_spitzer_2008}), while
planetesimals remain and continue planet formation processes.  In such
late phase of the disk, the sustained zone occupies larger volume of the
disk.
}

{
Although our nonlinear Ohm's law model is inspired by the 
lightning phenomena,
whether the Joule heating in this model takes the form of 
spatially and temporally concentrated stream of ionizing electrons --- lightning --- or not,
has yet to be studied in future works, employing nonthermal plasma studies.
We limit ourselves to pointing out a few distinguishing properties of such lightning which makes it an interesting subject.
First, the work-heat balance model suggests that in sustained zones the major portion of the shearing motion energy
is converted to lightning. This means that the lightning is one of the most dominant energy channel in 
the sustained zones. It will also pose a significant back-reaction to the accretion dynamics.
We will need to reconsider 
the contribution of lightning in situations it has been neglected
due to lack of energy, such as in chondrule formation
\citep{weidenschilling_production_1997}.
}

{
The second point is related to the redox environment the lightning creates.
Lightning induced by
collisional charging of
water ice dust overcomes the energetics problem
\citep{muranushi_dust-dust_2010}, but if applied as the chondrule heating source, 
it suffers from the redox environment mismatch.
Water vapor creates oxidizing environment
\citep{clayton_redox_1981,rubin_relationships_2005}
whereas major population of chondrules are considered to have formed in reducing environment
\citep{lofgren_dynamic_1989,connolly_carbon_1994,jones_chondrule_1997}.
However, the lightning in sustained zone proposed by \ISSS\ and studied in this paper, is 
a result of pure MHD process, and thus is redox-neutral. Therefore, it can potentially explain 
chondrule heating in both reducing and oxidizing environment.
}

\acknowledgments{
T.M. thanks Yuichiro Sekiguchi and Masaru Shibata for useful
discussions. We thank the referee for the valuable comments.
The simulations for this paper has been performed on
computer cluster at Kyoto University built by T.M, and
 TSUBAME2.0 Grid Cluster at Tokyo Institute of Technology.
The use of TSUBAME2.0  in this project was supported by
JHPCN through its program ``Joint Usage/Research Center for Interdisciplinary Large-scale Information Infrastructures.''
This work is supported by Grants-in-Aid from the Ministry of Education, Culture, Sports, Science, and Technology (MEXT) of Japan,
No. 24103506 (T.M.),
No. $22 \cdot 7006$ (S.O.),
 No.
18540238, No. 23244027, and No. 23103005 (S.I.).
}
%

\begin{landscape}
\begin{table}
\begin{tabular}{rc|ccc|ccccc}
runID&t&$\beta$&$\displaystyle \frac{\criticalCurrent}{\equipartitionCurrent}$&$\magneticReynolds$   &
$\displaystyle \left\langle \frac{ B^2}{8 \pi P_0}          \right\rangle$    &         
$\displaystyle \left\langle \frac{ -B_x B_y}{4 \pi P_0}     \right\rangle$    &         
$\displaystyle \left\langle \frac{\rho v_x \delta v_y}{P_0} \right\rangle $ &
$\displaystyle \left\langle \frac{J^2}{{\equipartitionCurrent}^2}  \right\rangle^{0.5} $  \\
\cline{1-9}
3160  &0&400  &$\infty$&$\infty$&$(1.99\pm0.60) \times {10}^{-1}$&$(1.05\pm0.37) \times {10}^{-1}$&$(3.01\pm1.49) \times {10}^{-2}$&$(2.04\pm0.24) \times {10}^{1}$\\
10    &0&800  &$\infty$&$\infty$&$(1.64\pm0.69) \times {10}^{-1}$&$(8.27\pm3.65) \times {10}^{-2}$&$(2.26\pm0.96) \times {10}^{-2}$&$(1.94\pm0.30) \times {10}^{1}$\\
220   &0&1600 &$\infty$&$\infty$&$(1.94\pm0.61) \times {10}^{-1}$&$(9.47\pm3.19) \times {10}^{-2}$&$(2.34\pm0.84) \times {10}^{-2}$&$(2.07\pm0.26) \times {10}^{1}$\\
3180  &0&3200 &$\infty$&$\infty$&$(1.30\pm0.27) \times {10}^{-1}$&$(6.08\pm1.47) \times {10}^{-2}$&$(1.68\pm0.52) \times {10}^{-2}$&$(1.85\pm0.18) \times {10}^{1}$\\

3612  &$0$&6400&$\infty$&$\infty$&$(7.16\pm2.57) \times {10}^{-2}$&$(3.14\pm0.85) \times {10}^{-2}$&$(8.28\pm2.51) \times {10}^{-3}$&$(1.51\pm0.15) \times {10}^{1}$\\
3616  &$0$&12800&$\infty$&$\infty$&$(3.61\pm0.83) \times {10}^{-2}$&$(1.70\pm0.35) \times {10}^{-2}$&$(6.15\pm2.40) \times {10}^{-3}$&$(1.24\pm0.10) \times {10}^{1}$\\
3620  &$0$&25600&$\infty$&$\infty$&$(3.61\pm0.84) \times {10}^{-2}$&$(1.63\pm0.36) \times {10}^{-2}$&$(5.40\pm1.84) \times {10}^{-3}$&$(1.22\pm0.09) \times {10}^{1}$\\

\\
\cline{1-9}
\\
3292  &$0$&400 &$1.0$ &$0.6$&$(2.92\pm1.56) \times {10}^{-1}$&$(1.67\pm1.04) \times {10}^{-1}$&$(4.81\pm3.63) \times {10}^{-2}$&$(2.10\pm0.43) \times {10}^{1}$&$\rceil$\\
3293  &$16\pi/\Omega$&400 &$1.0$ &$0.6$&$(2.74\pm1.43) \times {10}^{-1}$&$(1.62\pm0.99) \times {10}^{-1}$&$(4.63\pm3.05) \times {10}^{-2}$&$(2.06\pm0.48) \times {10}^{1}$&(a)\\
3294  &$18\pi/\Omega$&400 &$1.0$ &$0.6$&$(2.07\pm0.71) \times {10}^{-1}$&$(1.21\pm0.44) \times {10}^{-1}$&$(3.68\pm1.62) \times {10}^{-2}$&$(1.88\pm0.27) \times {10}^{1}$&$\bigcirc$\\
3295  &$20\pi/\Omega$&400 &$1.0$ &$0.6$&$(2.34\pm0.82) \times {10}^{-1}$&$(1.36\pm0.54) \times {10}^{-1}$&$(3.86\pm1.80) \times {10}^{-2}$&$(2.00\pm0.30) \times {10}^{1}$&$\rfloor$\\
\\
\cline{1-9}
\\
3352  &$0$&400 &$1.0$ &$0.2$&$(2.50\pm0.00) \times {10}^{-3}$&$(8.24\pm6.55) \times {10}^{-13}$&$(4.83\pm2.58) \times {10}^{-7}$&$(4.11\pm0.94) \times {10}^{-3}$&$\rceil$\\
3353  &$16\pi/\Omega$&400 &$1.0$ &$0.2$&$(2.03\pm1.27) \times {10}^{-1}$&$(1.27\pm0.86) \times {10}^{-1}$&$(4.07\pm2.99) \times {10}^{-2}$&$(1.50\pm0.60) \times {10}^{1}$&(b)\\
3354  &$18\pi/\Omega$&400 &$1.0$ &$0.2$&$(2.21\pm0.98) \times {10}^{-1}$&$(1.33\pm0.50) \times {10}^{-1}$&$(4.68\pm2.68) \times {10}^{-2}$&$(1.66\pm0.41) \times {10}^{1}$&$\bigtriangleup$\\
3355  &$20\pi/\Omega$&400 &$1.0$ &$0.2$&$(2.52\pm1.65) \times {10}^{-1}$&$(1.56\pm1.02) \times {10}^{-1}$&$(4.57\pm2.72) \times {10}^{-2}$&$(1.67\pm0.59) \times {10}^{1}$&$\rfloor$
\\
\\
\cline{1-9}
\\
3348  &$0$&400 &$10.0$&$0.2$&$(2.50\pm0.00) \times {10}^{-3}$&$(8.24\pm6.55) \times {10}^{-13}$&$(4.83\pm2.58) \times {10}^{-7}$&$(4.11\pm0.94) \times {10}^{-3}$&$\rceil$\\
3349  &$16\pi/\Omega$&400 &$10.0$&$0.2$&$(2.50\pm0.00) \times {10}^{-3}$&$(1.77\pm1.01) \times {10}^{-7}$&$(9.90\pm7.50) \times {10}^{-5}$&$(3.60\pm0.54) \times {10}^{-2}$&(c)\\
3350  &$18\pi/\Omega$&400 &$10.0$&$0.2$&$(2.50\pm0.00) \times {10}^{-3}$&$(2.31\pm1.64) \times {10}^{-6}$&$(6.06\pm9.16) \times {10}^{-5}$&$(3.49\pm0.54) \times {10}^{-2}$&$\times$\\
3351  &$20\pi/\Omega$&400 &$10.0$&$0.2$&$(2.50\pm0.00) \times {10}^{-3}$&$(4.85\pm3.58) \times {10}^{-7}$&$(6.62\pm8.64) \times {10}^{-5}$&$(3.43\pm0.36) \times {10}^{-2}$&$\rfloor$
\\

\end{tabular}
\caption{Statistics of the local simulations abridged. Each run is labeled by an integer.
  The re-start time is in
  the second column.
  Next three columns indicate the initial magnetic field strength,
  the critical current, and the magnetic Reynolds number.
  The
  physical quantity are represented in terms of the time average and
  standard deviation of the space average, i.e. $A$ is in the format
  $\overline{\langle A \rangle} \pm \left(\overline{{\langle A
      \rangle}^2} - {\overline{\langle A \rangle}}^2\right)^{0.5} $ . 
  In this table are runs for ideal MHD, runs in Figure \ref{figSMRIABC} 
  that represent the behavior in (a) active, (b) sustained, and (c) dead zones,
  runs that constitute sustained-dead zone boundaries for $\beta = 400, 3200$
}
\label{tableGrandStat1}
\end{table}

\end{landscape}

\begin{landscape}
\setcounter{table}{3}
\begin{table}
\begin{tabular}{cc|ccc|ccccc}
runID&t&$\beta$&$\displaystyle \frac{\criticalCurrent}{\equipartitionCurrent}$&$\magneticReynolds$   &
$\displaystyle \left\langle \frac{ B^2}{8 \pi P_0}          \right\rangle$  &         
$\displaystyle \left\langle \frac{ -B_x B_y}{4 \pi P_0}     \right\rangle$  &         
$\displaystyle \left\langle \frac{\rho v_x \delta v_y}{P_0} \right\rangle $ &
$\displaystyle \left\langle \frac{J^2}{{\equipartitionCurrent}^2}  \right\rangle^{0.5} $  \\
\cline{1-9}
3352  &0&400 &$1.0$ &$0.2$&$(2.50\pm0.00) \times {10}^{-3}$&$(8.24\pm6.55) \times {10}^{-13}$&$(4.83\pm2.58) \times {10}^{-7}$&$(4.11\pm0.94) \times {10}^{-3}$&\multirow{2}{*}{$\}\bigtriangleup$}\\
3353  &$16\pi/\Omega$&400 &$1.0$ &$0.2$&$(2.03\pm1.27) \times {10}^{-1}$&$(1.27\pm0.86) \times {10}^{-1}$&$(4.07\pm2.99) \times {10}^{-2}$&$(1.50\pm0.60) \times {10}^{1}$\\
3452  &0&400 &$0.9$ &$0.06$&$(2.50\pm0.00) \times {10}^{-3}$&$(3.05\pm2.55) \times {10}^{-13}$&$(4.29\pm2.47) \times {10}^{-7}$&$(2.22\pm0.52) \times {10}^{-3}$&\multirow{2}{*}{$\}\times$}\\
3453  &$16\pi/\Omega$&400 &$0.9$ &$0.06$&$(2.50\pm0.00) \times {10}^{-3}$&$(1.22\pm0.75) \times {10}^{-8}$&$(2.29\pm1.58) \times {10}^{-4}$&$(3.02\pm0.46) \times {10}^{-2}$\\
3448  &0&400 &$0.3$ &$0.06$&$(2.50\pm0.00) \times {10}^{-3}$&$(3.05\pm2.55) \times {10}^{-13}$&$(4.29\pm2.47) \times {10}^{-7}$&$(2.22\pm0.52) \times {10}^{-3}$&\multirow{2}{*}{$\}\bigtriangleup$}\\
3449  &$16\pi/\Omega$&400 &$0.3$ &$0.06$&$(2.29\pm2.53) \times {10}^{-1}$&$(1.43\pm1.71) \times {10}^{-1}$&$(4.61\pm4.95) \times {10}^{-2}$&$(1.39\pm0.90) \times {10}^{1}$\\
3436  &$0$&400 &$0.3$ &$0.02$&$(2.50\pm0.00) \times {10}^{-3}$&$(9.13\pm12.07) \times {10}^{-13}$&$(1.20\pm1.54) \times {10}^{-7}$&$(3.43\pm2.16) \times {10}^{-4}$&\multirow{2}{*}{$\}\times$}\\
3437  &$16\pi/\Omega$&400 &$0.3$ &$0.02$&$(2.50\pm0.00) \times {10}^{-3}$&$(2.39\pm1.08) \times {10}^{-9}$&$(8.02\pm6.85) \times {10}^{-5}$&$(1.17\pm0.23) \times {10}^{-2}$\\
3432  &$0$&400 &$0.1$ &$0.02$&$(2.50\pm0.00) \times {10}^{-3}$&$(9.55\pm12.38) \times {10}^{-13}$&$(8.32\pm8.82) \times {10}^{-8}$&$(2.91\pm1.54) \times {10}^{-4}$&\multirow{2}{*}{$\}\bigtriangleup$}\\
3433  &$16\pi/\Omega$&400 &$0.1$ &$0.02$&$(2.00\pm2.42) \times {10}^{-1}$&$(1.14\pm1.50) \times {10}^{-1}$&$(3.54\pm4.61) \times {10}^{-2}$&$(1.23\pm0.98) \times {10}^{1}$\\
\\
\cline{1-9}
\\
3372  &$0$&3200&$1.0$ &$0.2$&$(1.11\pm0.62) \times {10}^{-1}$&$(5.22\pm2.79) \times {10}^{-2}$&$(1.39\pm0.68) \times {10}^{-2}$&$(1.58\pm0.31) \times {10}^{1}$&\multirow{2}{*}{$\}\bigcirc$}\\
3373  &$16\pi/\Omega$&3200&$1.0$ &$0.2$&$(1.00\pm0.41) \times {10}^{-1}$&$(4.29\pm1.61) \times {10}^{-2}$&$(9.90\pm3.04) \times {10}^{-3}$&$(1.52\pm0.24) \times {10}^{1}$\\
3460  &$0$&3200&$0.9$ &$0.06$&$(3.12\pm0.00) \times {10}^{-4}$&$(3.02\pm1.86) \times {10}^{-12}$&$(4.68\pm2.21) \times {10}^{-7}$&$(1.89\pm0.39) \times {10}^{-3}$&\multirow{2}{*}{$\}\times$}\\
3461  &$16\pi/\Omega$&3200&$0.9$ &$0.06$&$(3.35\pm0.24) \times {10}^{-4}$&$(9.82\pm10.64) \times {10}^{-6}$&$(3.36\pm3.24) \times {10}^{-5}$&$(5.07\pm2.78) \times {10}^{-2}$\\
3456  &$0$&3200&$0.3$ &$0.06$&$(3.12\pm0.00) \times {10}^{-4}$&$(3.02\pm1.86) \times {10}^{-12}$&$(4.68\pm2.21) \times {10}^{-7}$&$(1.89\pm0.39) \times {10}^{-3}$&\multirow{2}{*}{$\}\bigtriangleup$}\\
3457  &$16\pi/\Omega$&3200&$0.3$ &$0.06$&$(1.19\pm0.48) \times {10}^{-1}$&$(5.44\pm1.88) \times {10}^{-2}$&$(1.42\pm0.51) \times {10}^{-2}$&$(1.65\pm0.26) \times {10}^{1}$\\
3444  &$0$&3200&$0.3$ &$0.02$&$(3.12\pm0.00) \times {10}^{-4}$&$(8.60\pm6.12) \times {10}^{-13}$&$(5.18\pm2.54) \times {10}^{-7}$&$(1.55\pm0.35) \times {10}^{-3}$&\multirow{2}{*}{$\}\bigtriangleup$}\\
3445  &$16\pi/\Omega$&3200&$0.3$ &$0.02$&$(9.38\pm4.06) \times {10}^{-2}$&$(4.50\pm1.88) \times {10}^{-2}$&$(1.15\pm0.46) \times {10}^{-2}$&$(1.29\pm0.27) \times {10}^{1}$\\
3440  &$0$&3200&$0.1$ &$0.02$&$(3.12\pm0.00) \times {10}^{-4}$&$(8.60\pm6.12) \times {10}^{-13}$&$(5.18\pm2.54) \times {10}^{-7}$&$(1.55\pm0.35) \times {10}^{-3}$&\multirow{2}{*}{$\}\bigtriangleup$}\\
3441  &$16\pi/\Omega$&3200&$0.1$ &$0.02$&$(8.15\pm3.91) \times {10}^{-2}$&$(3.79\pm1.70) \times {10}^{-2}$&$(1.05\pm0.50) \times {10}^{-2}$&$(1.41\pm0.23) \times {10}^{1}$\\

\end{tabular}
\caption{ (continued) }\label{tableGrandStat2}
\end{table}

\end{landscape}

\end{document}